\documentclass[aps,prd,preprint,groupedaddress]{revtex4-1}
\usepackage{graphicx}
\usepackage{amsmath}
\usepackage{bm}

\begin{document}
\begin{flushright}
  OU-HET-675 \ \\
\end{flushright}
\vspace{0mm}


\title{Gauge- and frame-independent  decomposition of nucleon spin}


\author{M.~Wakamatsu}
\email[]{wakamatu@phys.sci.osaka-u.ac.jp}
\affiliation{Department of Physics, Faculty of Science, \\
Osaka University, \\
Toyonaka, Osaka 560-0043, Japan}



\begin{abstract}
In a recent paper, we have shown that the way of gauge-invariant 
decomposition of the nucleon spin is not necessarily unique, but there still 
exists a preferable decomposition from the observational viewpoint. 
What was not complete in this argument is a fully satisfactory 
answer to the following questions. Does the proposed gauge-invariant
decomposition, especially the decomposition of the gluon total angular
momentum into its spin and orbital parts, correspond to observables
which can be extracted from high-energy deep-inelastic-scattering
measurements ? 
Is this decomposition not only gauge-invariant but also Lorentz frame-independent,
so that it is legitimately thought to reflect an intrinsic property of the nucleon ?
We show that we can answer both of these questions affirmatively, by making full
use of a gauge-invariant decomposition of covariant angular momentum tensor of
QCD in an arbitrary Lorentz frame.

\end{abstract}


\maketitle


\section{Introduction}

The so-called ``nucleon spin puzzle'' is still one of the most 
fundamental problems in hadron physics \cite{EMC88},\cite{EMC89}.
In the past few years, there have been several remarkable progresses
from the observational point of view. First, a lot of experimental
evidences have been accumulated, which indicate that the gluon
polarization inside the nucleon is likely to be
small \cite{COMPASSG06}\nocite{PHENIX06}\nocite{STAR06A}-\cite{STAR06B}.
At the least, it seems now widely accepted that the $U_A(1)$-anomaly
motivated explanation of the nucleon spin puzzle is disfavored.
Second, the quark spin fraction or the net longitudinal quark
polarization $\Delta q$ has been fairly precisely determined
through high-statistics measurements of the deuteron spin
structure function by COMPASS \cite{COMPASS05},\cite{COMPASS07} and
the HEREMES group \cite{HERMES07}.
According to these analyses, the portion of the
nucleon spin coming from the intrinsic quark spin is around
1/3. These observations necessarily attract a great deal of interest
in the role of orbital angular momenta of quark and gluon field
inside the nucleon.  


When one talks about the spin contents of the nucleon, however, one
cannot be unconcerned with the unsettled theoretical issues concerning
the decomposition of the nucleon spin. An especially difficult problem
here is the decomposition of the gluon total angular momentum into
its intrinsic spin and orbital parts.
Most people believe that the polarized gluon distribution function
is an observable quantity from high-energy
deep-inelastic-scattering (DIS) measurements \cite{Manohar90},\cite{CS82}.
On the other hand, it is often claimed that there is no gauge-invariant 
decomposition of the gluon total angular momentum into its spin and 
orbital parts \cite{JM90},\cite{Ji97PRL}.
Undoubtedly, this latter statement is closely connected
with another observation that there is no gauge-invariant local 
operator corresponding to the 1st moment of the polarized gluon 
distribution in the standard framework of operator-product expansion.
Since, the gauge principle is one of the most important principle of 
physics, which demands that only gauge-invariant quantities are 
measurable, how to reconcile these two conflicting observations is a 
fundamentally important problem in the physics of nucleon spin.

As the first step of the program, which aims at clearing up
the state of confusion, we have recently investigated the relationship
between the known decompositions of the nucleon
spin \cite{Wakamatsu10}. We showed that the gauge-invariant
decomposition advocated by Chen et al. \cite{Chen08},\cite{Chen09}
can be viewed as a nontrivial extension of the gauge-variant 
decomposition given by Jaffe and Manohar \cite{JM90}, so as to meet the 
gauge-invariance requirement of each term of the decomposition.
However, we have also pointed out that there is another gauge-invariant 
decomposition of the nucleon spin, which is closer to the Ji 
decomposition, while allowing the decomposition of the gluon total 
angular momentum into the spin and orbital parts.  After clarifying the
reason why the gauge-invariant decomposition of the nucleon spin is not
unique, we emphasized the possible superiority of our decomposition 
to that of Chen et al. on the ground of observability. To be 
more concrete, we developed an argument in favor of Ji's proposal to 
obtain a full decomposition of the nucleon spin \cite{Ji97}. It supports the 
widely-accepted experimental project, in which one first determines
the total angular momentum of quarks and gluons through 
generalized-parton-distribution (GPD) analyses and then extract the 
orbital angular momentum contributions of quarks and gluons by 
subtracting the intrinsic spin parts of quarks and gluons, which can 
be determined through polarized DIS measurements.
Unfortunately, our argument lacks a finishing touch in the respect that
we did not give a rigorous proof that the quark and gluon intrinsic
spin contributions in our gauge-invariant decomposition in fact
coincides with the quark and gluon polarizations extracted from the
polarized DIS analyses.
Another question, which is not unrelated to the above problem, is as 
follows. Since our gauge-invariant decomposition as well as that of 
Chen et al. are given in a specific Lorentz frame, we could not
give a definite answer to the question whether these decompositions
have a frame-independent meaning or not. The purpose of the present
paper is to solve these remaining problems.
We will show that these questions can be 
solved simultaneously, by making full use of a gauge-invariant
decomposition of covariant angular-momentum tensor of QCD in an
arbitrary Lorentz frame.

The plan of the paper is as follows. In sect.II, 
we show that we can make a gauge-invariant decomposition of the 
covariant angular-momentum tensor of QCD in an arbitrary Lorentz
frame, even without fixing gauge explicitly.
Next, in sect.III, the nucleon forward matrix element of the 
Pauli-Lubansky vector expressed in terms of the covariant
angular-momentum tensor and the nucleon momentum is utilized
to obtain a gauge- and frame-independent decomposition
 of the nucleon spin.
In sect.IV, we clarify the relation between our decomposition and
the high-energy DIS observables. Summary and conclusion of our
analyses are then given in sect.V.

\section{Gauge-invariant decomposition of covariant angular-momentum
tensor of QCD}

Following Jaffe and Manohar \cite{JM90}, we start with a Belinfante
symmetrized expression for 
QCD energy momentum tensor given by
\begin{equation}
 T^{\mu \nu} \ = \ T^{\mu \nu}_q \ + \ T^{\mu \nu}_g ,
\end{equation}
where
\begin{eqnarray}
 T^{\mu \nu}_q &=& \frac{1}{2} \,\bar{\psi} \,
 (\,\gamma^{\mu} \,i \,D^{\nu} \ + \  
 \gamma^{\nu} \,i \,D^{\mu} \,) \,\psi , \label{QCD-EM-tensor-q} \\
 T^{\mu \nu}_g &=& 2 \,\mbox{Tr} \,(\,F^{\mu \alpha} \,F_{\alpha}{}^\nu 
 \ - \ \frac{1}{4} \,g^{\mu \nu} \,F^2 \,) . \label{QCD-EM-tensor-g}
\end{eqnarray}
Here, $T^{\mu \nu}$ is conserved, $\partial_{\mu} T^{\mu \nu} = 0$, symmetric, 
$T^{\mu \nu} = T^{\nu \mu}$, and gauge invariant. The QCD angular momentum 
tensor $M^{\mu \nu \lambda}$ is a rank-3 tensor constructed from
$T^{\mu \nu}$ as 
\begin{equation}
 M^{\mu \nu \lambda} \ \equiv \ x^{\nu} \,T^{\mu \lambda} 
 \ - \ x^{\lambda} \,T^{\mu \nu} .
\end{equation}
$M^{\mu \nu \lambda}$ is conserved, $\partial_{\mu} M^{\mu \nu \lambda} = 0$,
and gauge-invariant, if $T^{\mu \nu}$ is symmetric and conserved.
Another noteworthy property of $M^{\mu \nu \lambda}$,
which was emphasized by Jaffe and Manohar, is that it has no totally
antisymmetric part, which means that it satisfies the identity
\begin{equation}
 \epsilon_{\alpha \mu \nu \lambda} \, M^{\mu \nu \lambda} \ = \ 0,
\end{equation}
or equivalently
\begin{equation}
 M^{\mu \nu \lambda} \ + \ M^{\lambda \mu \nu} \ + \ M^{\nu \lambda \mu} 
 \ = \ 0.
\end{equation}
As shown in \cite{JM90}, by using the identity
\begin{eqnarray}
 &\,& \bar{\psi} \,(\,x^{\nu} \,\gamma^{\lambda} \ - \ 
 x^{\lambda} \,\gamma^{\nu} \,) \,
 i \,D^{\mu} \,\psi \ - \ \bar{\psi} \,\gamma^{\mu} \, 
 (\,x^{\nu} \,i \,D^{\lambda} \ - \ x^{\lambda} \,i \,D^{\nu} \,) \,\psi 
 \nonumber \\
 &=& \epsilon^{\mu \nu \lambda \beta} \, 
 \bar{\psi} \,\gamma_{\beta} \,\gamma_5 \,\psi \ - \ 
 \frac{1}{2} \,\,\partial_{\alpha} \, 
 [\, (x^{\nu} \,\epsilon^{\mu \lambda \alpha  \beta} \ - \  
 x^{\lambda} \,\epsilon^{\mu \nu \alpha \beta} \,) \,
 \bar{\psi} \,\gamma_{\beta} \,\gamma_5 \,\psi \,] ,
\end{eqnarray}
the quark part of $M^{\mu \nu \lambda}$ can gauge-invariantly be decomposed in
the following way
\begin{eqnarray}
 M^{\mu \nu \lambda}_q &=& 
 \frac{1}{2} \,\epsilon^{\mu \nu \lambda \beta} \,\,  
 \bar{\psi} \,\gamma_{\beta} \,\gamma_5 \,\psi \ + \ 
 \bar{\psi} \,\gamma^{\mu} \,
 (\,x^{\nu} \,i \,D^{\lambda} \ - \ x^{\lambda} \,i \,D^{\nu} \,) \,\psi ,
\end{eqnarray}
{\it up to a surface term}.
In remarkable contrast, it is a wide-spread belief that the gluon part of 
$M^{\mu \nu \lambda}$ cannot be gauge-invariantly decomposed into the 
intrinsic spin and orbital angular momentum
contributions \cite{JM90},\cite{Ji97PRL}.
The gauge-invariant version of the decomposition of $M^{\mu \nu \lambda}$
given in the paper by Jaffe and Manohar is therefore given as
\begin{equation}
 M^{\mu \nu \lambda} \ = \ M^{\mu \nu \lambda}_q \ + \ 
 M^{\mu \nu \lambda}_g \ + \ \mbox{total divergence} ,
\end{equation}
with
\begin{eqnarray}
 M^{\mu \nu \lambda}_q &=& \frac{1}{2} \,\epsilon^{\mu \nu \lambda \beta} \, 
 \bar{\psi} \,\gamma_{\beta} \,\gamma_5 \,\psi \ + \ 
 \bar{\psi} \,\gamma^\mu \,(x^{\nu} \,i \,D^{\lambda}
 \ - \ x^{\lambda} \,i \,D^{\nu} \,) \,\psi ,\\
 M^{\mu \nu \lambda}_g &=& 2 \,\mbox{Tr} \,
 [\, x^{\nu} \,F^{\mu \alpha} \,F_{\alpha}{}^\lambda 
 \ - \ x^{\lambda} \,F^{\mu \alpha} \,F_{\alpha}{}^\nu \,] \ - \ 
 \frac{1}{2} \,\mbox{Tr} \,F^2 \,
 [\,x^{\nu} \,g^{\mu \lambda} \ - \ x^{\lambda} \,g^{\mu \nu} \,] .
\end{eqnarray}
Note that this is essentially the covariant version of the Ji
decomposition \cite{Ji97PRL}.
It should also be noted that the 2nd term of $M^{\mu \nu \lambda}_g$
contributes only to Lorentz boosts, so that it has nothing to do
with nucleon spin decomposition.

Somewhat surprisingly, however, basically by following the idea
proposal by Chen et al. \cite{Chen08},\cite{Chen09}, we can make
a gauge-invariant decomposition of $M^{\mu \nu \lambda}_g$, at least formally.
The idea is to decompose the gluon field into two parts as
\begin{equation}
 A^{\mu} \ = \ A^{\mu}_{phys} \ + \ A^{\mu}_{pure} ,
\end{equation}
with $A^{\mu}_{pure}$ a pure-gauge term transforming in the same way as
the full $A^{\mu}$ does, and always giving null field strength, and 
$A^{\mu}_{phys}$ a physical part of $A^{\mu}$ transforming in the same 
manner as $F^{\mu \nu}$ does, i.e. covariantly. 
That is, the two important properties of this decomposition is the 
condition  for the pure-gauge part of the field, 
\begin{equation}
 F^{\mu \nu}_{pure} \ \equiv \ \partial^{\mu} \,A^{\nu}_{pure} 
 \ - \ \partial^{\nu} \,A^{\mu}_{pure}
 \ - \ i \,g \,[\, A^{\mu}_{pure}, A^{\nu}_{pure} \,] \ = \ 0,
 \label{pure-gauge}
\end{equation}
and the gauge transformation properties of the two parts : 
\begin{eqnarray}
 A^{\lambda}_{phys} (x) 
 &\rightarrow& U (x) \,A^{\lambda}_{phys} (x) \,U^{-1} (x) , 
 \label {gtr-phys} \\ 
 A^{\lambda}_{pure} (x)
 &\rightarrow& U (x) \,\left(A^{\lambda}_{pure} (x) 
 \ + \ \frac{i}{g} \,\,\partial^{\lambda} \,\right) \,U^{-1} (x) .
 \label{gtr-pure}
\end{eqnarray}
As a matter of course, these conditions are not enough to uniquely 
fix gauge.
To uniquely fix gauge, Chen et al. proposed to impose some 
additional gauge-fixing condition, which is a generalization of the
Coulomb gauge condition in the case of QED.
(The detail of the gauge-fixing problem is discussed also in the recent
researches \cite{Cho10},\cite{Wong10}.)
Alternatively, one can take the light-cone gauge with some 
appropriate boundary condition for the gauge field.
In either case, these extra gauge-fixing procedure necessarily
breaks the Lorentz symmetry.
Fortunately, we find it possible to accomplish a gauge-invariant
decomposition of covariant rank-3 tensor $M^{\mu \nu \lambda}$
based on the above conditions (\ref{pure-gauge}) $\sim$ (\ref{gtr-pure})
only, while postponing a concrete gauge-fixing
procedure until the later stage.
The usefulness of such covariant formulation should become apparent if one
tries to compare the relation between the nucleon spin decomposition in 
different gauges and in different Lorentz frames. 

Now, we explain the derivation of a gauge-invariant decomposition of 
$M^{\mu \nu \lambda}$ in some detail, since this decomposition plays a 
central role in our following discussion.
First, by using the identity
\begin{eqnarray}
 F^{\alpha \lambda} &\equiv& \partial^{\alpha} \,A^{\lambda} 
 \ - \ \partial^{\lambda} \,A^{\alpha} \ - \ 
 i \,g \,[\, A^{\alpha}, A^{\lambda} \,] \ = \ 
 D^{\alpha} \,A^{\lambda} \ - \ \partial^{\lambda} \,A^{\alpha} ,
\end{eqnarray}
with $D^{\alpha} \equiv \partial^{\alpha} - i \,g [\, A^{\alpha}, \cdot \,]$ 
being the covariant derivative for the adjoint representation of
color $SU(3)$, one can 
easily prove the identity
\begin{eqnarray}
 x^{\nu} \,F^{\mu \alpha} \,F_{\alpha}{}^\lambda
 \ - \ x^{\lambda} \,F^{\mu \alpha} \,F_{\alpha}{}^\nu
 &=& F^{\mu \alpha} \,(\,x^{\nu} \,D_{\alpha} \,A^{\lambda}
 \ - \ x^{\lambda} \,D_{\alpha} \,A^{\nu} \,) \nonumber \\
 &-& F^{\mu \alpha} \,(\,x^{\nu} \,\partial^{\lambda} 
 \ - \ x^{\lambda} \,\partial^{\nu} \,) \,A_{\alpha} .
\end{eqnarray}
This gives
\begin{eqnarray}
 x^{\nu} \,F^{\mu \alpha} \,F_{\alpha}{}^\lambda 
 \ - \ x^{\lambda} \,F^{\mu \alpha} \,F_{\alpha}{}^\nu
 &=& F^{\mu \alpha} \,(\,x^{\nu} \,D_{\alpha} \,A^{\lambda}_{phys} 
 \ - \ x^{\lambda} \,D_{\alpha} \,A^{\nu}_{phys} \,) \nonumber \\
 &-& F^{\mu \alpha} \,(\,x^{\nu} \,\partial^{\lambda} 
 \ - \ x^{\lambda} \,\partial^{\nu} \,) \,A^{phys}_{\alpha} \nonumber \\
 &+& F^{\mu \alpha} \,(\,x^{\nu} \,D_{\alpha} \,A^{\lambda}_{pure}
 \ - \ x^{\lambda} \,D_{\alpha} \,A^{\nu}_{pure} \,) \nonumber \\
 &-& F^{\mu \alpha} \,(\,x^{\nu} \,\partial^{\lambda}
 \ - \ x^{\lambda} \,\partial^{\nu} \,) \,A^{pure}_{\alpha} .
 \label{fterm-decomp}
\end{eqnarray}
The sum of the 3rd and 4th terms can be transformed in the following way :
\begin{eqnarray}
 &\,& F^{\mu \alpha} \,[\, (\,x^{\nu} \,D_{\alpha} \,A^{\lambda}_{pure}
 \ - \ x^{\lambda} \,D_{\alpha} \,A^{\nu}_{pure} \,) \ - \ 
 (\,x^{\nu} \,\partial^{\lambda} 
 \ - \ x^{\lambda} \,\partial^{\nu} ) A^{pure}_{\alpha} \,] \nonumber \\
 &=& F^{\mu \alpha} \,[\, x^{\nu} \,(\, D_{\alpha} \,A^{\lambda}_{pure} 
 \ - \ \partial^{\lambda} \,A^{pure}_{\alpha} \,) \ - \ 
 x^{\lambda} \,(\,D_{\alpha} \,A^{\nu}_{pure}
 \ - \ \partial^{\nu} \,A_{\alpha}^{pure} ) \,] \nonumber \\
 &=& F^{\mu \alpha} \,\{\, x^{\nu} \,
 (\,\partial_{\alpha} \,A^{\lambda}_{pure} 
 \ - \ \partial^{\lambda} \,A_{\alpha}^{pure}
 \ - \ i \,g \,[\, A_{\alpha}^{pure}, A^{\lambda}_{pure} \,]
 \ - \ i \,g \,[\, A_{\alpha}^{phys}, A^{\lambda}_{pure} \,] \,) \nonumber \\
 &\,& \ \ \ \ 
 - \ x^{\lambda} \,(\, \partial_{\alpha} \,A^{\nu}_{pure}
 \ - \ \partial^{\nu} \,A_{\alpha}^{pure}
 \ - \ i \,g \,[\, A_{\alpha}^{pure}, A_{pure}^{\nu} \,] 
 \ - \ i \,g \,[\, A_{\alpha}^{phys}, A^{\nu}_{pure} \,] 
 \,) \,\} \nonumber \\
 &=& - \,i \,g \,F^{\mu \alpha} \, 
 (\,x^{\nu} \,[\, A_{\alpha}^{phys}, A_{pure}^{\lambda} \,] 
 \ - \ x^{\lambda} \,[\, A_{\alpha}^{phys}, A^{\nu}_{pure} \,] \,) .
\end{eqnarray}
Here, we have used the pure-gauge condition (\ref{pure-gauge}) for
the pure-gauge part of $A^{\mu}$.
Adding up the 2nd term of (\ref{fterm-decomp}) to the above sum, we
obtain
\begin{eqnarray}
 &-& \,F^{\mu \alpha} \,(\,x^{\nu} \,\partial^{\lambda} \ - \ 
 x^{\lambda} \partial^{\nu} \,) \,A_{\alpha}^{phys} 
 \ - \ i \,g \,F^{\mu \alpha} 
 \,(\, x^{\nu} \, [\, A_{\alpha}^{phys}, A_{pure}^{\lambda} \,] 
 \ - \ x^{\lambda} \,
 [\, A_{\alpha}^{phys}, A_{pure}^{\nu} \,] \,) \nonumber \\
 &=& - \,F^{\mu \alpha} \,\{\, x^{\nu} \,
 (\, \partial^{\lambda} \ - \ i \,g \, 
 [\, A_{pure}^{\lambda}, A_{\alpha}^{phys} \,] \,) \ - \ 
 x^{\lambda} \,(\,\partial^{\nu}
 \ - \ i \,g \,[\, A_{pure}^{\nu}, A_{\alpha}^{phys} \,] \,) \} \nonumber \\
 &=& F^{\mu \alpha} \,(\,x^{\nu} \,D_{pure}^{\lambda} \,
 A_{\alpha}^{phys} \ - \ x^{\lambda} \,D_{pure}^{\nu} \,A_{\alpha}^{phys} \,).
\end{eqnarray}
Here, we have introduced the {\it pure-gauge covariant derivative} by
\begin{equation}
 D_{pure}^{\lambda} \ \equiv \ \partial^{\lambda} 
 \ - \ i \,g \,[\, A_{pure}^{\lambda}, \,\cdot \,\,] .
\end{equation}
As a consequence of the manipulation above, we obtain a fairly simple
relation : 
\begin{eqnarray}
 x^{\nu} \,F^{\mu \alpha} \,F_{\alpha}{}^{\lambda}
 \ - \ x^{\lambda} \,F^{\mu \alpha} \,F_{\alpha}{}^{\nu}
 &=& F^{\mu \alpha} \,(\,x^{\nu} \,D_{\alpha} \,A_{phys}^{\lambda}
 \ - \ x^{\lambda} \,D_{\alpha} \,A_{phys}^{\nu} \,) \nonumber \\
 &-& F^{\mu \alpha} \,(\, x^{\nu} \,D_{pure}^{\lambda} \,A_{\alpha}^{phys}
 \ - \ x^{\lambda} \,D_{pure}^{\nu} A_{\alpha}^{phys} \,) .
 \label{fterm-decomp1}
\end{eqnarray} 
Now, making use of the relation $D_{\alpha} F^{\alpha \mu} = 
\partial_{\alpha} F^{\alpha \mu} - i \,g \,[ A_{\alpha}, F^{\alpha \mu} \,]$,
it is straightforward to prove the identity : 
\begin{eqnarray}
 \partial_{\alpha} \,\mbox{Tr} \,
 (\,F^{\alpha \mu} \,x^{\nu} \,A^{\lambda}
 \ - \ F^{\alpha \mu} \,x^{\lambda} \,A^{\nu} \,)
 &=& \mbox{Tr} \,\{\, (\,D_{\alpha} \,F^{\alpha \mu} \,) \,
 (\, x^{\nu} \,A^{\lambda} \ - \ x^{\lambda} \,A^{\nu} \,) \nonumber \\
 &-& F^{\mu \alpha} \,(\,x^{\nu} \,D_{\alpha} \,A^{\lambda}
 \ - \ x^{\lambda} \,D_{\alpha} \,A^{\nu} \,) \nonumber \\
 &+& F^{\mu \lambda} \,A^{\nu} \ - \ F^{\mu \nu} \,A^{\lambda} \,\} .
 \label{fterm-decomp2}
\end{eqnarray}
It is also obvious from the above derivation that a similar identity holds
even though we replace the
fields $A^{\lambda}$ and $A^{\nu}$ above by their physical parts, i.e.
$A_{phys}^{\lambda}$ and $A_{phys}^{\nu}$ :
\begin{eqnarray}
 \partial_{\alpha} \,\mbox{Tr} \,
 (\, F^{\alpha \mu} \,x^{\nu} \,A_{phys}^{\lambda}
 \ - \ F^{\alpha \mu} \,x^{\lambda} \,A_{phys}^{\nu} \,)
 &=& \mbox{Tr} \,\{\, (\,D_{\alpha} \,F^{\alpha \mu} \,) \,
 (\,x^{\nu} \,A_{phys}^{\lambda} \ - \ x^{\lambda} \,A_{phys}^{\nu} \,)
 \nonumber \\
 &-& F^{\mu \alpha} \,(\, x^{\nu} \,D_{\alpha} \,A_{phys}^{\lambda}
 \ - \ x^{\lambda} \,D_{\alpha} \,A_{phys}^{\nu} \,) \nonumber \\
 &+& F^{\mu \lambda} \,A_{phys}^{\nu} 
 \ - \ F^{\mu \nu} \,A_{phys}^{\lambda} \,\} . \label{fterm-deriv}
\end{eqnarray}
Combining (\ref{fterm-decomp1}) and (\ref{fterm-deriv}), we thus find
the relation
\begin{eqnarray}
 &\,& \mbox{Tr} \,(\,x^{\nu} \,F^{\mu \alpha} \,F_{\alpha}{}^{\lambda}
 \ - \ x^{\lambda} \,F^{\mu \alpha} \,F_{\alpha}{}^{\nu} \,) \ + \ 
 \partial_{\alpha} \,\mbox{Tr} \,
 (\,F^{\alpha \mu} \,x^{\nu} \,A_{phys}^{\lambda}
 \ - \ F^{\alpha \mu} \,x^{\lambda} \,A_{phys}^{\nu} \,) \ \ \ \ \ 
 \nonumber \\
 &=& \mbox{Tr} \,\{\,(\,D_{\alpha} \,F^{\alpha \mu} \,) \,
 (\,x^{\nu} \,A_{phys}^{\lambda} \ - \ x^{\lambda} \,A_{phys}^{\nu} \,)
 \ - \ 
 F^{\mu \alpha} \,(\,x^{\nu} \,D_{pure}^{\lambda}
 \ - \ x^{\lambda} \,D_{pure}^{\nu} \,) \,A_{\alpha}^{phys} \nonumber \\
 &\,& \hspace{70mm}
 + \ \,\,F^{\mu \lambda} \,A_{phys}^{\nu} \ - \ F^{\mu \nu} \,A_{phys}^{\nu} \,
 \,\} . \label{identity-g}
\end{eqnarray}
After all these steps, we eventually arrive at the following
decomposition for the QCD angular momentum tensor (we call it the
decomposition (I)) : 
\begin{eqnarray}
 M^{\mu \nu \lambda} &=& M^{\mu \nu \lambda}_{q - spin}
 \ + \ M^{\mu \nu \lambda}_{q - OAM} \ + \ M^{\mu \nu \lambda}_{g - spin}
 \ + \ M^{\mu \nu \lambda}_{g - OAM} \nonumber \\
 &+& \ M^{\mu\nu \lambda}_{boost} \ + \ \mbox{total divergence} ,
 \label{decomposition1}
\end{eqnarray}
where
\begin{eqnarray}
 M^{\mu \nu \lambda}_{q - spin}
 &=& \frac{1}{2} \,\epsilon^{\mu \nu \lambda \sigma} \,\bar{\psi} 
 \,\gamma_{\sigma} \,\gamma_5 \,\psi , \label{decomposition1A} \\
 M^{\mu \nu \lambda}_{q - OAM}
 &=& \bar{\psi} \,\gamma^\mu \,(\,x^{\nu} \,i \,D^{\lambda} 
 \ - \ x^{\lambda} \,i \,D^{\nu} \,) \,\psi \label{decomposition1B} \\
 M^{\mu \nu \lambda}_{g - spin} 
 &=& 2 \,\mbox{Tr} \,[\, F^{\mu \lambda} \,A^{\nu}_{phys} 
 \ - \ F^{\mu \nu} \,A^{\lambda}_{phys} \,], \label{decomposition1C} \\
 M^{\mu \nu \lambda}_{g - OAM} 
 &=& - \,2 \,\mbox{Tr} \,[\, F^{\mu \alpha} \,
 (\,x^{\nu} \,D^{\lambda}_{pure}
 \ - \ x^{\lambda} \,D^{\nu}_{pure} \,) \,A_{\alpha}^{phys} \,],
 \nonumber \\
 &\,& + \, 2 \,\mbox{Tr} \,[\, (\,D_{\alpha} \,F^{\alpha \mu} \,)
 \,(\,x^{\nu} \,A^{\lambda}_{phys} \ - \ 
 x^{\lambda} \,A^{\nu}_{phys} \,) \,], \label{decomposition1D} \\
 M^{\mu \nu \lambda}_{boost} 
 &=& - \,\frac{1}{2} \,\mbox{Tr} \,F^2 \,(\,x^{\nu} \,g^{\mu \lambda} 
 \ - \ x^{\lambda} \,g^{\mu \nu} \,) . \label{decomposition1E}
\end{eqnarray}
In the above decomposition, $M^{\mu \nu \lambda}_{q-spin}$ and
$M^{\mu \nu \lambda}_{q-OAM}$ respectively correspond to the spin
and orbital angular momentum parts of quarks, while
$M^{\mu \nu \lambda}_{g-spin}$ and $M^{\mu \nu \lambda}_{g-OAM}$ to
the spin and orbital angular momentum parts of gluons.
(At the quantum level, there is some delicacy in the identification
of the term $M^{\mu \nu \lambda}_{q-spin}$ with the 
intrinsic quark spin part. This will be discussed in the next section.)
We have already pointed out that the term $M^{\mu \nu \lambda}_{boost}$
contributes only to the Lorentz boots.
An important feature of the above decomposition
(\ref{decomposition1}) of $M^{\mu \nu \lambda}$ 
is that each piece is separately gauge invariant. Since this is already
obvious for the quark part, let us confirm it below for less trivial gluon 
part.

The gauge invariance of the $M^{\mu \nu \lambda}_{g-spin}$ and 
the 2nd term of $M^{\mu \nu \lambda}_{g-OAM}$ can easily be
convinced if one remembers the covariant transformation property
(\ref{gtr-phys}) of the physical part of $A^\mu$
as well as the covariant transformation
property of the field strength tensor $F^{\mu \nu}$.
Less trivial is the 1st term of gluon orbital part $M^{\mu \nu \lambda}_{g-OAM}$.
We first notice that, under a gauge
transformation, $D^\lambda_{pure} \,A^{phys}_\alpha$ transform as
\begin{eqnarray}
 D^\lambda_{pure} \,A^{phys}_\alpha &\equiv&
 \partial^\lambda \,A^{phys}_\alpha \ + \ i \,g \,
 [\,A^\lambda_{pure}, A^{phys}_\alpha \,] \nonumber \\
 &\rightarrow& \partial^\lambda \,(\,U \,A^{phys}_\alpha \,U^{-1} \,)
 \ - \ i \,g \,[\,U \,
 (\,A^\lambda_{pure} \ + \ \frac{i}{g} \,\partial^\lambda \,) \,U^{-1},
 U \,A^{phys}_\alpha \,U^{-1} \,] \nonumber \\
 &=& U \,(\,\partial^\lambda \,A^{phys}_\alpha \ - \ i \,g \,
 [\,A^\lambda_{pure}, A^{phys}_\alpha \,] \,) \,U^{-1} \nonumber \\
 &=& U \,D^\lambda_{pure} \,A^{phys}_\alpha \,U^{-1} .
\end{eqnarray}
This means that $D^\lambda_{pure} \,A^{phys}_\alpha$ transforms
{\it covariantly} under a gauge transformation. The gauge-invariance of the
1st term of $M^{\mu \nu \lambda}_{g-OAM}$ should be almost obvious from
this fact. Altogether, this confirms the fact that each term of the
decomposition (I) is in fact separately gauge-invariant.

Note that the gluon orbital angular momentum contribution
$M^{\mu \nu \lambda}_{g-OAM}$ consists of two terms.
Using the QCD equation of motion
\begin{equation}
 \left(\,D^\mu \,F_{\mu \nu} \,\right)^a \ = \ - \,g \,
 \bar{\psi} \,\gamma_\nu \,T^a \,\psi ,
\end{equation}
the 1st term of $M^{\mu \nu \lambda}_{g-OAM}$ can also be expressed in the
form,
\begin{eqnarray}
 2 \,\mbox{Tr} \,\left\{\, (\,D_\alpha \,F^{\alpha \mu} \,) \,
 (\,x^\nu \,A^\lambda_{phys} \ - \ x^\lambda \,A^\nu_{phys} \,) \,\right\}
 \ = \ - \,g \,\bar{\psi} \,\gamma^\mu \,(\,x^\nu \,A^\lambda_{phys}
 \ - \ x^\lambda \,A^\nu_{phys} \,) \,\psi .
\end{eqnarray}
Undoubtedly, this term is a covariant generalization of the ``potential
angular momentum'' {\it a la} Konopinski \cite{Konopinski78} as pointed
out in our previous paper \cite{Wakamatsu10}.
Since this term is {\it solely} gauge-invariant, one has
a freedom to combine it with another gauge-invariant term, for example,
with the quark orbital angular momentum term of the
decomposition (I). This leads to another gauge invariant decomposition
of $M^{\mu \nu \lambda}$ given as (this will be called the decomposition (II))
\begin{eqnarray}
 M^{\prime \mu \nu \lambda} &=& M^{\prime \mu \nu \lambda}_{q - spin}
 \ + \ M^{\prime \mu \nu \lambda}_{q - OAM} \ + \ 
 M^{\prime \mu \nu \lambda}_{g - spin}
 \ + \ M^{\prime \mu \nu \lambda}_{g - OAM} \nonumber \\
 &+& \ M^{\prime \mu \nu \lambda}_{boost} \ + \ \mbox{total divergence} ,
 \label{decomposition2}
\end{eqnarray}
where
\begin{eqnarray}
 M^{\prime \mu \nu \lambda}_{q - spin}
 &=& \frac{1}{2} \,\epsilon^{\mu \nu \lambda \sigma} \,\bar{\psi} 
 \,\gamma_{\sigma} \,\gamma_5 \,\psi , \\
 M^{\prime \mu \nu \lambda}_{q - OAM}
 &=& \bar{\psi} \,\gamma^\mu \,(\,x^{\nu} \,i \,D^{\lambda}_{pure} 
 \ - \ x^{\lambda} \,i \,D^{\nu}_{pure} \,) \,\psi \\
 M^{\prime \mu \nu \lambda}_{g - spin} 
 &=& 2 \,\mbox{Tr} \,[\, F^{\mu \lambda} \,A^{\nu}_{phys} 
 \ - \ F^{\mu \nu} \,A^{\lambda}_{phys} \,], \\
 M^{\prime \mu \nu \lambda}_{g - OAM} 
 &=& - \,2 \,\mbox{Tr} \,[\, F^{\mu \alpha} \,
 (\,x^{\nu} \,D^{\lambda}_{pure}
 \ - \ x^{\lambda} \,D^{\nu}_{pure} \,) \,A_{\alpha}^{phys} \,], \\
 M^{\prime \mu \nu \lambda}_{boost} 
 &=& - \,\frac{1}{2} \,\mbox{Tr} \,F^2 \,(\,x^{\nu} \,g^{\mu \lambda} 
 \ - \ x^{\lambda} \,g^{\mu \nu} \,) .
\end{eqnarray}
Noteworthy here is the fact that the intrinsic spin parts are just the
common in the two decompositions (I) and (II) for both of quarks and gluons, i.e.
\begin{eqnarray}
 M^{\prime \mu \nu \lambda}_{q - spin}
 &=& M^{\mu \nu \lambda}_{q - spin}, \\
 M^{\prime \mu \nu \lambda}_{g - spin}
 &=& M^{\mu \nu \lambda}_{g - spin},
\end{eqnarray}
whereas the orbital parts are critically different for both of quarks and gluons, i.e.
\begin{eqnarray}
 M^{\prime \mu \nu \lambda}_{q - OAM}
 &\neq& M^{\mu \nu \lambda}_{q - OAM}, \\
 M^{\prime \mu \nu \lambda}_{g - OAM}
 &\neq& M^{\mu \nu \lambda}_{g - OAM},
\end{eqnarray}
although it holds that the sum of the quark and gluon orbital angular momenta
precisely coincides in the two decompositions, i.e.
\begin{equation}
 M^{\prime \mu \nu \lambda}_{q - OAM} \ + \ 
 M^{\prime \mu \nu \lambda}_{g - OAM} \ = \ 
 M^{\mu \nu \lambda}_{q - OAM} \ + \ 
 M^{\mu \nu \lambda}_{g - OAM}.
\end{equation}
One might think that the decomposition (II) can be thought of as a
covariant generalization of the gauge-invariant decomposition of
Chen et al.~\cite{Chen08},\cite{Chen09}. 
Actually, gauge is not definitely fixed yet in our treatment.
We still have complete freedom to choose any desired gauge
compatible with the decomposition of the gluon field into its
physical and pure-gauge parts.
By choosing a ``generalized Coulomb gauge'' advocated by Chen et al.
in a suitable Lorentz frame, the above decomposition would in fact
reduce to that of Chen et al. On the other hand, if one takes the
light-cone gauge with some residual gauge degrees of freedom,
the decomposition (II)
reproduces the gauge-invariant decomposition of the nucleon spin proposed by
Bashinsky and Jaffe \cite{BJ99}, which was  proposed on the basis of the
light-cone-gauge formulation of parton distribution functions.
 (For confirmation of this statement above, see the discussion in sect.4.)
On the other hand, we already know that the Chen decomposition reduces to
the Jaffe-Manohar decomposition after a particular gauge fixing.
Then, the above argument altogether indicates that the known three
decompostions, i.e. those of Jaffe and Manohar, of Bashinsky and Jaffe, and
of Chen et al. are all contained in our decomposition (II) so that gauge-equivalent.
In other words, they are the same decomposition from the physical viewpoint.

We have pointed out that, in the two decompositions (I) and (II) of the
angular-momentum tensor, the difference exists only in the orbital parts.
Here, let us look into simpler quark part more closely. What appears in our
decomposition (I) is a covariant generalization
of the so-called ``dynamical'' or ``mechanical''
orbital angular momentum of quarks. On the other hand, what appears
in the decomposition (II) is a nontrivial gauge-invariant extension
of ``canonical'' orbital angular momentum.
This difference is of crucial physical significance, since, as emphasized in
our previous paper \cite{Wakamatsu10},
the dynamical orbital angular momentum is a measurable quantity,
whereas the canonical one is not.
In fact, the common knowledge of standard electrodynamics tells us
that the momentum appearing in the equation of motion
with the Lorentz force is the so-called dynamical momentum
$\bm{\Pi} \ = \ \bm{p} - q \,\bm{A}$ with the full gauge field, not
the canonical momentum $\bm{p}$ or its nontrivial extension
$\bm{p} - q \,\bm{A}_{pure}$. To convince it,
let us consider the motion of a charged particle with mass $m$
and a charge $e$ ($e < 0$ for the electron) under the influence of
static electric and magnetic field given as \cite{BookSakurai95}
\begin{equation}
 \bm{E} \ = \ - \,\nabla \,\phi, \ \ \ \ 
 \bm{B} \ = \ \nabla \times \bm{A} .
\end{equation}
The hamiltonian, which describes the motion of the charged particle, is
given by
\begin{equation}
 H \ = \ \frac{\bm{\Pi}^2}{2 \,m} \ + \ e \,\phi ,
\end{equation}
with
\begin{equation}
 \bm{\Pi} \ \equiv \ \bm{p} \ - \ e \,\bm{A} .
\end{equation}
The equation of motion for this charged particle becomes
\begin{equation}
 m \,\frac{d^2 \bm{x}}{d t^2} \ = \ \frac{d \bm{\Pi}}{d t} \ = \ 
 e \,\left[\,\bm{E} \ + \ \frac{1}{2} \,\left(\,
 \frac{d \bm{x}}{d t} \times \bm{B} \ - \ \bm{B} \times \frac{d \bm{x}}{d t}
 \,\right) \,\right] .
\end{equation}
This equation of motion dictates that the momentum accompanying the mass
flow of a charged particle is the {\it dynamical} momentum
$\bm{\Pi} = \bm{p} - e \,\bm{A}$ containing the full gauge field $\bm{A}$,
not the {\it canonical} momentum $\bm{p}$ or its nontrivial extension
$\bm{p} - e \,\bm{A}_{pure}$. Similarly, the angular momentum accompanying
the mass flow of a charge particle is the dynamical orbital angular momentum
$\bm{x} \times \bm{\Pi} = \bm{x} \times (\bm{p} - e \,\bm{A})$, not
$\bm{x} \times \bm{p}$ or $\bm{x} \times (\bm{p} - e \,\bm{A}_{pure})$.

In the subsequent sections, we try to make the above statement on the
observability of our decomposition more concrete first for the quark part.
The analysis is then extended to the gluon part to accomplish
a complete decomposition of the nucleon spin.

\section{Frame-independence of our nucleon spin decomposition}

Our discussion in this section is based on our recommendable decomposition (I)
of the QCD angular momentum tensor $M^{\mu \nu \lambda}$
given in (\ref{decomposition1}) - (\ref{decomposition1E}).
The nucleon spin sum rule is obtained by evaluating the forward matrix 
element of the tensor $M^{012}$ in the equal-time quantization, or 
that of the tensor $M^{+12}$ in the light-cone quantization. This gives the normalization condition
\begin{equation}
 \langle P, s \,|\, M^{012} \,|\, P, s \rangle \,/ \,
 \langle P, s \,|\, P, s \rangle \ = \ \frac{1}{2} ,
\end{equation}
in the equal-time quantization, or
\begin{equation}
 \langle P,s \,|\, M^{+12} \,|\, P, s \rangle \,/ \,
 \langle P,s \,|\, P, s \rangle \ = \ \frac{1}{2} ,
\end{equation}
in the light-cone quantization.
Here, $|P,s \rangle$ stands for a plane-wave nucleon state with
momentum $P_{\mu}$ and spin $s_{\mu}$.
An alternative method to obtain the nucleon spin sum rule is to evaluate the forward matrix element of the helicity operator \cite{Ji97PRD}
\begin{equation}
 W^{\mu} s_\mu \ = \ \bm{J} \cdot \hat{\bm{P}} \ = \ 
 \frac{\bm{J} \cdot \bm{P}}{|\bm{P}|},
\end{equation}
where
\begin{equation}
 W^{\mu} \ = \ - \,\frac{1}{2 \,\sqrt{P^2}} \,\, 
 \epsilon^{\mu \alpha \beta \gamma} \,J_{\alpha \beta} \,P_{\gamma}
\end{equation}
with $J^{\alpha \beta} = M^{0 \alpha \beta}$, is the Pauli-Lubansky 
vector \cite{Lubanski42}, while $P_{\mu}$ and $s_{\mu}$ are the momentum
and the spin vector of the nucleon satisfying the relations : 
\begin{equation}
 P^2 \ = \ M^2, \ \ s^2 \ = \ - \,1, \ \ P \cdot s \ = \ 0 .
\end{equation}
The normalization condition in this case is 
\begin{equation}
 \langle P,s \,|\, W^{\mu} s_{\mu} \,|\, P,s \rangle \,/ \,
 \langle P,s \,|\, P,s \rangle \ = \ \frac{1}{2} .
\end{equation}

In either case, for spin decomposition of the nucleon, we need to know
forward matrix element of each term of the r.h.s. of
(\ref{decomposition1}). We first consider the forward matrix element of
$M^{\mu \nu \lambda}_{q - spin}$.
Although we have naively called this term the intrinsic quark 
spin contribution to $M^{\mu \nu \lambda}$, there is some delicacy.
As first recognized by Jaffe and Manohar \cite{JM90}, and later
elaborated in \cite{SW00} and \cite{BLT04},
$\bar{\psi} \,\gamma_{\sigma} \, \gamma_5 \, \psi = A_{\sigma}^{(0)}$
is the flavor-singlet axial current
and it enters $M^{\mu \nu \lambda}$ in the form 
$\frac{1}{2} \,\epsilon^{\mu \nu \lambda \sigma} A_{\sigma}^{(0)}$. 
However, Jaffe and Manohar also noticed the fact that $M^{\mu \nu \lambda}$
should have no totally antisymmetric part.
This observation, combined with the fact that the 
total derivative term has no forward matrix element, leads to the 
conclusion that the forward matrix element of $M^{\mu \nu \lambda}$ 
cannot have a term proportional to $\epsilon^{\mu \nu \lambda \sigma}$.
This means that the term of this form coming from
$M^{\mu \nu \lambda}_{q-spin} = \frac{1}{2} \,
\epsilon^{\mu \nu \lambda \sigma} \,A_\sigma^{(0)}$
must exactly be canceled by a 
similar term coming from the ``orbital piece'' of $M^{\mu \nu \lambda}$.
First, we shall verify this fact explicitly for the quark part of 
$M^{\mu \nu \lambda}$.
Later, we will show that a similar situation occurs
also for the gluon  part.
In general, the forward matrix element of 
$M^{\mu \nu \lambda}_{q-spin}$ is specified by the flavor-singlet 
axial charge $a^{(0)}_q$ as 
\begin{equation}
 \langle P,s \,|\, M^{\mu \nu \lambda}_{q - spin} (0) \,|\, P, s \rangle 
 \ = \ M \,a^{(0)}_q \,\,\epsilon^{\mu \nu \lambda \sigma} \,s_{\sigma} .
\end{equation}
It is a widely-known fact that, at the quantum level, an ambiguity 
arises, due to the $U_A (1)$ anomaly of QCD, concerning the relation 
between the flavor-singlet axial charge and the net quark polarization 
$\Delta q$ (or the net contribution of the intrinsic quark spin to the 
nucleon spin). In the most popular factorization (or renormalization) 
scheme, i.e. in the $\overline{\rm MS}$ scheme, $a^{(0)}_q$ can just be
identified with $\Delta q$.
On the other hand, there is another class of renormalization scheme 
called the Adler-Bardeen (AB) schemes, in which $a^{(0)}_q$ is given by 
$a^{(0)}_q = \Delta q - 2 \,n_f \,(\alpha_s / 4 \pi) \,\Delta g$ with
$\Delta g$ the net gluon polarization, and $n_f$ the number
of quark flavors.
An advantage of the AB scheme is that $\Delta q$ is completely 
scale-independent. Nonetheless, there is no compelling reason to stick 
to this scheme. Without any loss of generality, we can choose the 
$\overline{\rm MS}$ scheme, in which the forward matrix element of 
$M^{\mu \nu\lambda}_{q - spin}$ gives the net quark spin contribution 
to the nucleon spin through the previously-mentioned sum rule.

Next, we investigate the forward matrix element of the quark orbital angular
momentum part $M^{\mu \nu \lambda}_{q - OAM}$.
This part of the current takes a general form of 
\begin{equation}
 M^{\mu \nu \lambda} (x) \ = \ x^{\nu} \,O^{\mu \lambda} (x)
 \ - \ x^{\lambda} \,O^{\mu \nu} (x) ,
\end{equation}
so that the evaluation of its forward matrix element needs some care. 
The method is well-known and given by the following limiting
procedure \cite{JM90} : 
\begin{eqnarray}
 &\,& \langle P,s \,|\, M^{\mu \nu \lambda} (0) \,|\,P,s \rangle
 \ = \ 
 \lim_{\Delta \rightarrow 0} \,\,i \,\,\frac{\partial}{\partial \Delta}_\nu
 \left\langle P + \frac{\Delta}{2} , s \,\right| \,
 O^{\mu \lambda} (0) \,\left| \, P - \frac{\Delta}{2},s \right\rangle
 \ - \ (\nu \ \leftrightarrow \ \lambda) . \ \ \ \ \ \ \ 
 \label{limiting}
\end{eqnarray}
(More sound formulation of this limiting procedure with use of
wave packets instead of plane waves was later elaborated in
\cite{BLT04} and \cite{SW00}.)
To make use of the above formula, we first note that 
$M^{\mu \nu \lambda}_{q-OAM}$ can be expressed as
\begin{equation}
 M^{\mu \nu \lambda}_{q-OAM} 
 \ = \ x^{\nu} \,O^{\mu \lambda}_2 \ - \ x^{\lambda} \,O^{\mu \nu}_2 .
\end{equation}
with 
\begin{equation}
 O^{\mu \nu}_2 \ = \ \bar{\psi} \,\gamma^\mu \,i \,D^{\nu} \,\psi .
\end{equation}
It is important to recognize that this rank-2 tensor $O^{\mu \nu}_2$ 
entering $M^{\mu \nu \lambda}_{q-OAM}$ is different from the quark part 
of the QCD energy-momentum tensor
\begin{equation}
 T^{\mu \nu}_q 
 \ = \ \frac{1}{2} \,\bar{\psi} \,\gamma^{ \{ \mu } \,i \,D^{\nu \}} \,\psi .
\end{equation}
by the effect of symmetrization. (Here we use the notation 
$a^{ \{\mu} b^{\nu \} } = a^{\mu} b^{\nu} + a^{\nu} b^{\mu}$ 
and $a^{ [ \mu} b^{\nu ] } = a^{\mu} b^{\nu} - a^{\nu} b^{\mu}$.)
Then, while the nonforward matrix element of $T^{\mu \nu}_q$ is
characterized by three form factors as 
\begin{eqnarray}
 \left\langle P + \frac{\Delta}{2} , s \,\right|\, 
 T^{\mu \nu}_q (0) \,\left|\, P - \frac{\Delta}{2},s \right\rangle
 &=& A_q (\Delta^2) \,P^{\mu} \,P^{\nu}
 \ + \ \frac{B_q (\Delta^2)}{2 M} \,
 P^{ \{ \mu} \epsilon^{\nu \} \alpha \beta \sigma} \,
 s_{\alpha} \,P_{\beta} \,i \,\Delta_{\sigma} \nonumber \\
 &+& C_q (\Delta^2) \,M^2 \,g^{\mu \nu} \ + \ O (\Delta^2) ,
 \label{tensor-qcd}
\end{eqnarray}
the nonforward matrix element of $O^{\mu \nu}_2$ can contain extra terms
which are antisymmetric in $\mu$ and $\nu$ as 
\begin{eqnarray}
 \left\langle P + \frac{\Delta}{2}, s \,\right|\, 
 O^{\mu \nu}_2 (0) \,\left| \, P - \frac{\Delta}{2},s \right\rangle
 &=& A_q (\Delta^2) P^{\mu} \,P^{\nu}
 \ + \ \frac{B_q (\Delta^2)}{2 M} \,P^{ \{ \mu} \, 
 \epsilon^{\nu \} \,\alpha \,\beta \sigma} \,s_{\alpha} \,P_{\beta}
 \,\,i \,\Delta_{\sigma} \nonumber \\
 &+& \frac{\tilde{B_q} (\Delta^2)}{2 M} \,P^{[ \mu} \epsilon^{\nu ] 
 \alpha \beta \sigma} \,s_{\alpha} \,P_{\beta} \,\,i \,\Delta_{\sigma}
 \ + \ M \,D_q (\Delta^2) \,\epsilon^{\mu \nu \alpha \beta} \,
 s_{\alpha} \,i \,\Delta_{\beta} \nonumber \\
 &+& C_q (\Delta^2) \,M^2 \,g^{\mu \nu} \ + \ O (\Delta^2) .
 \label{tensor-o2}
\end{eqnarray}
(The above parametrizations of the nucleon matrix elements of rank-2
tensors were criticized in the paper by Bakker, Leader, and
Trueman \cite{BLT04}. They argue that, if $T^{\mu \nu}$ transforms as
a second-rank tensor, its nonforward matrix elements do not transform
covariantly. Only by first factoring out the wave functions, i.e. the
Dirac spinors in the case of nucleon matrix elements, the relevant
function sandwiched by the initial and final wave functions transform
covariantly. Nevertheless, they themselves confirmed that, despite
this problem of the parametrization of the nucleon nonforward matrix
elements, the treatment of Jaffe and Manohar give just the correct
answer at least for the longitudinal spin sum rule of the nucleon,
which is of our current interest. For the sake of simplicity, we therefore
follow the treatment of Jaffe and Manohar at the cost of complete
stringency.)

Now, a key observation of our nucleon spin decomposition is as follows.
As shown by Shore and White \cite{SW00}, the two rank-2
tesnsors $T^{\mu \nu}_q$ and $O^{\mu \nu}_2$ are not completely
independent. They are related through the following identity : 
\begin{eqnarray}
 x^{\nu} \,T^{\mu \lambda}_q \ - \ x^{\lambda} \,T^{\mu \nu}_q
 &=& x^{\nu} \,O^{\mu \lambda}_2 \ - \ x^{\lambda} \,O^{\mu \nu}_2
 \nonumber \\
 &+& \frac{1}{2} \,\epsilon^{\mu \nu \lambda \sigma} \, 
 \bar{\psi} \,\gamma_{\sigma} \,\gamma_5 \,\psi \ + \ 
 \mbox{total divergence} .
\end{eqnarray}
By evaluating the forward matrix element of this identity,
one can prove that all the form factors, appearing in (\ref{tensor-qcd})
and (\ref{tensor-o2}), are not independent but obey the following
relation : 
\begin{equation}
 \tilde{B}_q (0) \ = \ 0, \ \ \ 2 \,D_q (0) \ = \ a^{(0)}_q .
\end{equation}
As a consequence, we find that the forward matrix element 
of $M_{q - OAM}^{\mu \nu \lambda}$ is given by
\begin{eqnarray}
 \left\langle P, s \,\right|\, 
 M^{\mu \nu \lambda}_{q - OAM} (0) \,
 \left|\, P, s \right\rangle
 &=& \frac{B_q (0)}{2 M} \,P^{ \{ \mu} \epsilon^{\lambda \} \nu
 \alpha \beta} \,P_{\alpha} \,s_{\beta} \ - \ 
 (\,\nu \leftrightarrow \lambda \,) \nonumber \\
 &-& M \,a^{(0)}_q \,\epsilon^{\mu \nu \lambda \sigma} \,s_{\sigma} .
 \label{q-oam}
\end{eqnarray}
As emphasized in \cite{JM90} and explicitly shown in \cite{SW00}, 
the axial-charge term, which is totally, 
antisymmetric in the indices $\mu, \nu, \lambda,$ cancels in the 
forward matrix elements of 
$M^{\mu \nu \lambda}_{q - spin}$ plus $M^{\mu \nu \lambda}_{q - OAM}$ 
to give
\begin{eqnarray}
 \left\langle P, s \,\right| 
 M^{\mu \nu \lambda}_{q-spin} (0) \, + \, 
 M^{\mu \nu \lambda}_{q - OAM} (0) 
 \left|\, P, s \right\rangle
 \, = \, \frac{B_q (0)}{2 M} \,P^{\{ \mu} \epsilon^{\nu \} \alpha \beta} 
 \,P_{\alpha} \,s_{\beta} \ - \ ( \nu \leftrightarrow \lambda) . \ \ \ \
 \label{q-j} 
\end{eqnarray}  
It can be shown that $B_q (0)$ just coincide with the total angular 
momentum $J_q$ carried by the quark fields, 
\begin{equation}
 B_q (0) \ = \ J_q .
\end{equation} 
%

Now we turn to the discussion of much more difficult
gluon part. Despite a lot of efforts, whether the total gluon angular
momentum $J_g$ can be gauge-invariantly decomposed into the spin and
orbital parts is still a controversial problem.
That it is possible at the formal level has
been shown in a series of paper by Chen et al. \cite{Chen08},\cite{Chen09}
and has been confirmed in our recent paper \cite{Wakamatsu10}.
However, these decompositions were achieved in a 
particular Lorentz frame. What we are looking for here is a Lorentz covariant formulation.
An advantage of Lorentz covariant formulation
is that we can make clear
the relation between the nucleon spin decompositions obtained in
different Lorentz frames. Furthermore, as we shall see shortly,
it also turns out to reveal an important physics, which was masked
in a noncovariant formulation.
We first look into the forward matrix element of our gluon-spin operator
\begin{equation}
 M^{\mu \nu \lambda}_{g - spin} 
 \ = \ 2 \,\mbox{Tr} \,[\, F^{\mu \lambda} \,A^{\nu}_{phys} 
 \ - \ F^{\mu \nu} \,A^{\lambda}_{phys} \,]  \ = \ 
 2 \,\mbox{Tr} \,[\, F^{\mu \lambda} \,A^{\nu}_{phys} 
 \ + \ F^{\nu \mu} \,A^{\lambda}_{phys} \,] .
\end{equation}
We first emphasize that this operator is gauge-invariant, so that it is delicately
different from the gauge-variant current
\begin{equation}
 M^{\mu \nu \lambda}_{(g)} (spin) 
 \ \equiv \ 2 \,\mbox{Tr} \,[\, F^{\mu \nu} \,A^{\nu} \ + \ 
 F^{\nu \mu} \,A^{\lambda} \,] ,
\end{equation}
which was naively identified with the gluon spin operator in the paper
by Jaffe and Manohar \cite{JM90}.
In the same paper, however, they pointed out a very
interesting fact. According to them, the analogy with the quark part
would have led to expect $M^{\mu \nu \lambda}_{(g)} (spin)$ to be
\begin{eqnarray}
 \epsilon^{\mu \nu \lambda \sigma} \,K_{\sigma} 
 &=& 2 \,\mbox{Tr} \,[\, F^{\nu \lambda} \,A^{\mu} 
 \ + \ A^{\nu} \,F^{\lambda \mu}
 \ + \ A^{\lambda} \,F^{\mu \nu} \,] \ + \ 
 2 \,i \,g \,\mbox{Tr} \,A^{\mu} \,[\, A^{\nu}, A^{\lambda} \,] ,
\end{eqnarray}
which is totally antisymmetric in the three indices $\mu,\nu,\lambda$.
Here
\begin{eqnarray}
 k_\mu &\equiv& \frac{\alpha_S}{2 \,\pi} \,K_\mu \ = \ 
 \frac{\alpha_S}{2 \,\pi} \,\epsilon_{\mu \nu \alpha \beta} \,
 \mbox{Tr} \,A^\nu \,\left[\,F^{\alpha \beta} \ - \ \frac{2}{3} \,
 A^\alpha \,A^\beta \,\right] .
\end{eqnarray}
is the gauge-variant Chern-Simons current, whose divergence is
related to the well-known topological charge density of QCD as
\begin{equation}
 \partial^\mu \,k_\mu \ = \ \frac{\alpha_S}{2 \,\pi} \,
 \mbox{Tr} \,F^{\mu \nu} \,\tilde{F}_{\mu \nu} .
\end{equation}
Owing to the symmetry difference, 
$\epsilon^{\mu \nu \lambda \sigma} K_{\sigma}$ 
and $K^{\mu \nu \lambda}_{(g)} (spin)$ are not in the same 
representation of the Lorentz group  \cite{JM90}. The former belongs to 
$(\frac{1}{2}, \frac{1}{2})$, while the latter contains 
$(\frac{1}{2}, \frac{3}{2}) \oplus (\frac{3}{2}, \frac{1}{2})$ 
in addition to $(\frac{1}{2}, \frac{1}{2})$.
Historically, several authors advocated to use the forward matrix
element of the topological current to define the gluon axial charge
$a_g^{(0)} (0)$ or the gluon polarization
$\Delta g$ \cite{AR88},\cite{CCM88},\cite{ET88}.
(See also reviews \cite{AEL95},\cite{LR00}.)
However, some authors soon recognized that the gauge-variant nature
of the topological current $k_{\mu}$ prevents this
attempt \cite{Manohar91},\cite{BB91},\cite{SW00}.
The argument goes as follows.
The nonforward matrix element of the topological current 
$k^{\mu}$ is characterized by two form factors as
\begin{equation}
 \left\langle P + \frac{\Delta}{2}, s \,\right|\, k^{\mu} \,
 \left|\, P - \frac{\Delta}{2},s \right\rangle
 \ = \ 2 \,M \,s^{\mu} \,a_g^{(0)} (\Delta^2)
 \ + \ \Delta^{\mu} \,(\Delta \cdot s) \,p_g (\Delta^2)  \ + \ 
 O (\Delta^2).
\end{equation}
Naively thinking, the 2nd term of the above equation would vanish in the
forward limit $\Delta^{\mu} \rightarrow 0$, so that one might expect that
\begin{equation}
 \langle P,s \,|\, k^{\mu} \,|\, P,s \rangle
 \ = \ 2 \,M \,s^{\mu} \,a_g^{(0)} (0)
\end{equation}
with the identification 
$a_g^{(0)} (0) = \frac{\alpha s}{4 \pi} \,\Delta g$.
However, it was soon recognized that the gauge-variant current $k^{\mu}$
couples to an unphysical Goldstone mode and the form factor
$p_g (\Delta^2)$ has a massless pole \cite{Manohar91},\cite{BB91}.
The structure of this pole depends on the adopted
gauge. It turns out that the forward matrix element of the topological current
is singular in general gauges. Although the matrix element is finite in the 
generalized axial gauges, $n \cdot A = 0$, its value still depends on 
the ways of taking the forward limit $\Delta \rightarrow 0$ so that
it is indefinite.
 
Now, we go back to our gauge-invariant operator 
$M^{\mu \nu \lambda}_{g - spin}$. It is instructive to rewrite 
$M^{\mu \nu \lambda}_{g - spin}$ in the form that contains the topological current
in itself as
\begin{eqnarray}
 M^{\mu \nu \lambda}_{g - spin} 
 &=& \epsilon^{\mu \nu \lambda \sigma} \,K_{\sigma} \nonumber \\
 &-& 2 \,\mbox{Tr} \,\{\, (\,F^{\lambda \nu} 
 \ + \ i \,g \,[\, A^{\lambda}, A^{\nu} \,] \,) \,A^{\mu} \,\} \nonumber \\
 &-& 2 \,\mbox{Tr} \,\{\, F^{\mu \lambda} \,A^{\nu}_{pure} 
 \ + \ F^{\nu \mu} \,A^{\lambda}_{pure} \,\} . \label{tensor-g-spin}
\end{eqnarray}
One might think that this manipulation is a little artificial.
Note, however, that it resembles the operation in the 
quark part, in which totally antisymmetric part 
$\frac{1}{2} \epsilon^{\mu \nu \lambda \sigma} 
\bar{\psi} \gamma_{\sigma} \gamma_5 \psi$ is separated from the total 
quark contribution $M^{\mu \nu \lambda}_q$.
An important difference with the quark case is that each term of
(\ref{tensor-g-spin}) is not separately gauge-invariant.
Nonetheless, the l.h.s of (\ref{tensor-g-spin}) is gauge-invariant by
construction, so that it is logically obvious that the 
gauge-dependencies of the three terms in the r.h.s. should exactly be
canceled. The argument above then indicates that the nonforward matrix
element of $M^{\mu \nu \lambda}_{g-spin}$ can be specified by
gauge-independent three form factors as
\begin{eqnarray}
 \left\langle P + \frac{\Delta}{2}, s \,\right|\, 
 M^{\mu \nu \lambda}_{g - spin} (0) \,
 \left|\, P - \frac{\Delta}{2}, s \right\rangle
 &=& 2 \,M \,\left(\frac{\alpha_s}{4 \,\pi} \right)^{-1} \,
 a_g^{(0)} (\Delta^2) \,\epsilon^{\mu \nu \lambda \sigma} \,
 s_{\sigma} \nonumber \\
 &+& v_g (\Delta^2) \,\epsilon^{\mu \nu \lambda \sigma} \, 
 \Delta_{\sigma} (\Delta \cdot s) \nonumber \\
 &+& w_g (\Delta^2) \,\Delta^{\mu} \,(\,\Delta^{\lambda} \,s^{\nu}
 \ - \ \Delta^{\nu} \,s^{\lambda} \,) \ + \ O (\Delta^2) . \ \ \ 
\end{eqnarray}
Now, an important difference with the past argument is that, since 
$M^{\mu \nu \lambda}_{g - spin}$ is manifestly gauge-invariant, there should
be no massless pole in either of the form factors 
$v_g (\Delta^2)$ and $w_g (\Delta^2)$. 
This means that the terms containing $v_g (\Delta^2)$ and $w_g (\Delta^2)$
vanish in the forward limit and the forward matrix element of
$M^{\mu \nu \lambda}_{g-spin}$ is unambiguously given by  
\begin{eqnarray}
 \langle P,s \,|\, M^{\mu \nu \lambda}_{g - spin} (0) \,|\, P,s \rangle
 &=& 2 \,M \,\left(\frac{\alpha_s}{4 \,\pi} \right)^{-1} \,a^{(0)}_g (0) \,\,
 \epsilon^{\mu \nu \lambda \sigma} \,s_{\sigma} \nonumber \\
 &=& 2 \,M \,\Delta g \,\,\epsilon^{\mu \nu \lambda \sigma} \,s_{\sigma} .
 \label{g-spin}
\end{eqnarray}
In short, although our gluon-spin operator is not necessarily
totally antisymmetric in the indices $\mu, \nu$ and $\lambda$,
only the totally antisymmetric part survives in its forward matrix
element. Although this seems somewhat mysterious, it certainly is a
consequence of logical reasoning explained above.

Our remaining task now is to evaluate the forward matrix element of 
$M^{\mu \nu \lambda}_{g - OAM}$. We first remember the fact that 
$M^{\mu \nu \lambda}_{g - OAM}$ can be expressed in the form
\begin{equation}
 M^{\mu \nu \lambda}_{g - OAM} 
 \ = \ x^{\nu} \,O^{\mu \lambda}_5 \ - \ x^{\lambda} \,O^{\mu \nu}_5 ,
\end{equation}
with 
\begin{eqnarray}
 O^{\mu \nu}_5 &=& - \,2 \,\mbox{Tr} \, 
 [\, F^{\mu \alpha} \,D^{\nu}_{pure} \,A^{phys}_{\alpha} \,] \ + \ 
 2 \,\mbox{Tr} \,[\,(\,D_{\alpha} \,F^{\alpha \mu} \,) \,A^{\nu}_{phys} \,] .
\end{eqnarray}
This should be compared with the net gluon  contribution to 
$M^{\mu \nu \lambda}$, which can be expressed as 
\begin{equation}
 M^{\mu \nu \lambda}_g 
 \ = \ x^{\nu} \,T^{\mu \lambda}_g \ - \ x^{\lambda} \,T^{\mu \nu}_g ,
\end{equation}
where $T^{\mu \nu}_g$ is the gluon contribution to the symmetric QCD
energy momentum tensor given by (\ref{QCD-EM-tensor-g}). 
There is a simple relation between $M^{\mu \nu \lambda}_g$ and
$M^{\mu \nu \lambda}_{g - OAM}$, however.   
That is, as is clear from (\ref{identity-g}), aside from the boost term,
$M^{\mu \nu \lambda}_g$
is different from the sum of 
$M^{\mu \nu \lambda}_{g-spin}$ and $M^{\mu \nu \lambda}_{g - OAM}$ 
only by a total divergence as 
\begin{equation}
 M^{\mu \nu \lambda}_g \ - \ \mbox{boost} \ = \ M^{\mu \nu \lambda}_{g - spin}
 \ + \ M^{\mu \nu \lambda}_{g - OAM} \ + \ \mbox{total divergence} .
 \label{diff-deriv}
\end{equation}
Note that this is a {\it key relation} in our gauge-invariant decomposition
of the gluon total angular momentum into its spin and orbital parts.

Now, we can proceed just in the same way as in the quark part.
The nonforward matrix element of 
$T^{\mu \nu}_g (0)$ and $O^{\mu\nu}_5 (0)$ are parametrized as  
\begin{eqnarray}
 \left\langle P + \frac{\Delta}{2}, s \,\right|\, 
 T^{\mu \nu}_g (0) \,\left|\, P - \frac{\Delta}{2},s \right\rangle
 &=& A_g (\Delta^2) \,P^{\mu} \,P^{\nu} 
 \ + \ \frac{B_g (\Delta^2)}{2 \,M} \,P^{ \{ \mu} 
 \epsilon^{\nu \} \alpha \beta \sigma} \,s_{\alpha} \, 
 P_{\beta} \,i \,\Delta_{\sigma} \nonumber \\
 &+& C_g (\Delta^2) \,M^2 \,g^{\mu \nu} \ + \ O (\Delta^2),
 \label{g-all} 
\end{eqnarray}
and
\begin{eqnarray}
 \left\langle P + \frac{\Delta}{2}, s \,\right|\, 
 O^{\mu \nu}_5 (0) \,\left|\, P - \frac{\Delta}{2},s \right\rangle
 &=& A_g (\Delta^2) \,P^{\mu} \,P^{\nu}
 \ + \ \frac{B_g (\Delta^2)}{2 \,M} \,P^{\{ \mu} \, 
 \epsilon^{\nu \} \alpha \beta \sigma} \, 
 s_{\alpha} \,P_{\beta} \,\,i \,\Delta_{\sigma} \nonumber \\
 &+& \frac{\tilde{B}_g (\Delta^2)}{2 \,M} \,
 P^{ [ \mu} \epsilon^{\nu ] \alpha \beta \sigma} \,
 s_{\alpha} \,P_{\beta} \,\,i \,\Delta_{\sigma}
 \ + \ M \,D_g (\Delta^2) \,
 \epsilon^{\mu \nu \lambda \sigma} 
 \,\,i \,\Delta_\lambda \,s_{\sigma} \nonumber \\
 &+& C_g (\Delta^2) \,M^2 \,g^{\mu \nu} \ + \ O (\Delta^2) .
 \label{g-oam}
\end{eqnarray}
By using the limiting procedure (\ref{limiting}), we thus have in the
forward limit : 
\begin{eqnarray}
 \langle P,s \,|\, M^{\mu \nu \lambda}_g (0) \,|\, P,s \rangle
 \ = \ \frac{B_g (0)}{2 \,M} \,P^{ \{ \mu} 
 \epsilon^{\lambda \} \nu \alpha \beta} \,s_{\alpha} \,P_{\beta}
 \ - \ (\,\nu \leftrightarrow \lambda \,) , \label{g-all2}
\end{eqnarray}  
and 
\begin{eqnarray}
 \langle P,s \,|\, M^{\mu \nu \lambda}_{g - OAM} (0) \,|\, P,s \rangle
 &=& \frac{B_g (0)}{2 \,M} \,P^{ \{ \mu} 
 \epsilon^{\lambda \} \nu \alpha \beta} \,s_{\alpha} \,P_{\beta}
 \ - \ (\,\nu \leftrightarrow \lambda \,) \nonumber \\
 &+& \frac{\tilde{B}_g (0)}{2 \,M} \,P^{ [ \mu} 
 \epsilon^{\lambda ] \nu \alpha \beta} \,s_{\alpha} \,P_{\beta}
 \ - \ (\,\nu \leftrightarrow \lambda \,) \nonumber \\
 &-& 2 \,M \,D_g (0) \,\epsilon^{\mu \nu \lambda \sigma} \,s_{\sigma} ,
 \label{g-oam2}
\end{eqnarray}
while we recall that
\begin{equation}
 \langle P, s \,| \,M^{\mu \nu \lambda}_{g-spin} (0) \,|\,P, s \rangle
 \ = \ 2 \,M \,\epsilon^{\mu \nu \lambda \sigma} \,s_\sigma \,\Delta g .
 \label{g-spin2}
\end{equation}
Then, in consideration of the fact that the total divergence term 
does not contribute to the forward matrix element, the relation
(\ref{diff-deriv}) together with (\ref{g-all2}),
(\ref{g-oam2}), (\ref{g-spin2}),  demands that
\begin{equation}
 D_g (0) \ = \ \Delta g, \ \ \ \tilde{B}_g (0) \ = \ 0 .
\end{equation}
We are then led to the desired result 
\begin{eqnarray}
 \langle P,s \,|\, M^{\mu \nu \lambda}_{g - spin} (0) \,|\, P,s \rangle 
 &=& 2 \,M \,\Delta g \,\epsilon^{\mu \nu \lambda \sigma} \,s_{\sigma} ,\\
 \langle P,s \,|\, M^{\mu \nu \lambda}_{g - OAM} (0) \,|\, P,s \rangle 
 &=& \frac{B_g(0)}{2 \,M} \,P^{ \{ \mu} 
 \epsilon^{\lambda \} \nu \alpha \beta} \,s_{\alpha} \,P_{\beta} 
 \ - \ (\,\nu \leftrightarrow \lambda \,) \nonumber \\
 &-& 2 \,M \,\Delta g \,\epsilon^{\mu \nu \lambda \sigma} \,s_{\sigma} .
\end{eqnarray}
which gives a gauge-invariant decomposition of $J_g$ into the spin 
and orbital parts. Again, the totally antisymmetric terms in the indices
$\mu, \nu, \lambda$ cancel in the forward matrix element of the sum of
$M^{\mu \nu \lambda}_{g-spin}$ and $M^{\mu \nu \lambda}_{g-OAM}$
to give
\begin{eqnarray}
 \left\langle P, s \,\right| 
 M^{\mu \nu \lambda}_{g-spin} (0) \, + \, 
 M^{\mu \nu \lambda}_{g-OAM} (0) 
 \left|\, P, s \right\rangle
 \, = \, \frac{B_g (0)}{2 M} \,P^{\{ \mu} \epsilon^{\nu \} \alpha \beta} 
 \,P_{\alpha} \,s_{\beta} \, - \, ( \nu \leftrightarrow \lambda) . \ \ \ \
 \label{g-j} 
\end{eqnarray}  

Let us summarize at this point what we have found.
We found that 
\begin{eqnarray} 
 \langle P, s \,|\,M^{\mu \nu \lambda} \,| \,P, s \rangle &=& 
 \langle P, s \,|\,M^{\mu \nu \lambda}_{q-spin} \,|\,P, s \rangle \ + \  
 \langle P, s \,|\,M^{\mu \nu \lambda}_{q-OAM} \,|\,P, s \rangle
 \nonumber \\
 &+& 
 \langle P, s \,|\,M^{\mu \nu \lambda}_{g-spin} \,|\,P, s \rangle \ + \ 
 \langle P, s \,|\,M^{\mu \nu \lambda}_{g-OAM} \,|\,P, s \rangle 
 \nonumber \\
 &+& \  
 \mbox{boost} .
\end{eqnarray}
with
\begin{eqnarray}
 \langle P,s \,|\, M^{\mu \nu \lambda}_{q - spin} (0) \,|\, P,s \rangle 
 &=&  M \,\Delta q \,\epsilon^{\mu \nu \lambda \sigma} \,s_{\sigma}, \\
 \langle P,s \,|\, M^{\mu \nu \lambda}_{q - OAM} (0) \,|\, P,s \rangle 
 &=& \frac{B_q(0)}{2 \,M} \,P^{ \{ \mu} \,
 \epsilon^{\lambda \} \nu \alpha \beta} \,s_{\alpha} \,P_{\beta} 
 \ - \ (\,\nu \leftrightarrow \lambda \,) \nonumber \\
 &-&  M \,\Delta q \,\epsilon^{\mu \nu \lambda \sigma} \,s_{\sigma}, \\
 \langle P,s \,|\, M^{\mu \nu \lambda}_{g - spin} (0) \,|\, P,s \rangle 
 &=& 2 \,M \,\Delta g \,\epsilon^{\mu \nu \lambda \sigma} \,s_{\sigma}, \\
 \langle P,s \,|\, M^{\mu \nu \lambda}_{g - OAM} (0) \,|\, P,s \rangle 
 &=& \frac{B_g(0)}{2 \,M} \,P^{ \{ \mu} \, 
 \epsilon^{\lambda \} \nu \alpha \beta} \,s_{\alpha} \,P_{\beta} \, 
 \ - \ (\,\nu \leftrightarrow \lambda \,) \nonumber \\
 &-& 2 \,M \,\Delta g \,\epsilon^{\mu \nu \lambda \sigma} \,s_{\sigma} .
\end{eqnarray}
We emphasize again that this is a completely gauge-invariant 
decomposition.
Inserting the above decomposition into the equation 
$\langle P,s \,|\, W^{\mu} \,s_{\mu} \,| P,s \rangle \,/ \, \langle  P,s 
\,|\, P,s \rangle = 1/2$ \cite{Ji97PRD}, one gets
\begin{equation}
 \frac{1}{2} \ = \ S_q \ + \ L_q \ + \ S_g \ + \ L_g  \ = \ J_q \ + \ J_g,
 \label{Nspin-SR}
\end{equation}
with
\begin{eqnarray}
 S_q &=& \frac{1}{2} \,\,\Delta q, \label{Nspin-SR1} \\
 L_q &=& B_q (0) \ - \ \frac{1}{2} \,\,\Delta q, \label{Nspin-SR2} \\
 S_g &=& \Delta g, \label{Nspin-SR3} \\
 L_g &=& B_g (0) \ - \ \Delta g . \label{Nspin-SR4}
\end{eqnarray}
This means that the individual contributions to the spin of the nucleon is
invariant under wide class of Lorentz transformation that preserve 
the helicity of the nucleon.
In this sense, we are now able to say that our
decomposition of the nucleon spin is not only gauge-invariant but
also basically Lorentz-frame independent.
A remaining important question is therefore as follows.
Can we give any convincing argument to show the observability of the
above decomposition ? A central task here is to verify whether the
above gluon spin term $S_g$ can in fact be identified with the
1st moment of the polarized gluon distribution determined by
high-energy polarized DIS analyses.
We try to answer this question in the next section.

\section{Observability of our nucleon spin decomposition}

It is a widely known fact that the quark and gluon total angular
momenta, i.e. $J_q$ and $J_g$, can in principle be extracted from
generalized-parton-distribution (GPD) analyses \cite{Ji97PRL}\cite{Ji97}.
Let us first confirm that our decomposition is compatible with
this common wisdom. Here, we closely follow the analysis by Shore
and White \cite{SW00}. We start with the standard definition of
unpolarized GPDs for quark and gluons given as
\begin{eqnarray}
 f_q (x,\xi,t) &=& \int \,\frac{d z^-}{2 \,\pi} \,
 e^{\,i \,\left( x + \frac{\xi}{2} \right) \,P^+ \,z^-} \nonumber \\
 &\,& \times \ \left\langle P + \frac{1}{2} \,\Delta \,| \,
 \bar{\psi}(0) \,\gamma^+ \,{\cal L}_g (0, z^-) \,\psi(z^-) \,|\,
 P - \frac{1}{2} \,\Delta \right\rangle , \nonumber \\
 \label{q-gpd}
 x \,P^+ \,f_g (x,\xi,t) &=& \int \,\frac{d z^-}{2 \,\pi} \,
 e^{\,i \,\left( x + \frac{\xi}{2} \right) \,P^+ \,z^-} \nonumber \\
 &\,& \times \ \left\langle P + \frac{1}{2} \,\Delta \,| \,
 2 \,\mbox{Tr} \,[\,F^{+ \alpha} (0) \,
 {\cal L}_g (0, z^-) \,F_\alpha^+ (z^-) \,|\,
 P - \frac{1}{2} \,\Delta \right\rangle ,
 \label{g-gpd} 
\end{eqnarray}
where $t = \Delta^2$, while ${\cal L}_g (a,b) = P \,
e^{\,- \,i \,g \,\int_b^a \,A \cdot ds}$ is the standard gauge link.
It is an easy exercise to derive the following 2nd moment sum rules
for $f_q (x, \xi, t)$ and $f_g (x, \xi, t)$ : 
\begin{eqnarray}
 \int_{-1}^1 \,x \,f_q (x, \xi, t) \,dx &=& 
 \left \langle P + \frac{\Delta}{2} \,| \,\bar{\psi}(0) \,
 \gamma^+ \,D^+ \,\psi (0) \,| \,P - \frac{\Delta}{2} \right\rangle
 \,/ \,(P^+)^2 , \\
 \label{q-1mom}
 \int_{-1}^1 \,x \,f_g (x, \xi, t) \,dx &=& 
 \left \langle P + \frac{\Delta}{2} \,| \, 
 2 \,\mbox{Tr} \,[\,F^{+ \alpha}(0) \,F_\alpha^+ (0) \,]
 \,| \,P - \frac{\Delta}{2} \right\rangle
 \,/ \,(P^+)^2 .
 \label{g-1mom}
\end{eqnarray}
The operators appearing in the r.h.s. of (\ref{q-1mom}) and
(\ref{g-1mom}) are respectively the $++$-component of the quark
and gluon parts of the QCD energy momentum tensor.
Especially simple here is the forward limit $t \rightarrow 0,\,
\xi \rightarrow 0$. In this limit, $f_q (x,\xi,t)$ and $f_g (x,\xi,t)$
reduce to the standard parton distribution functions (PDFs) of quarks and
gluons, i.e. $f_q (x)$ and $f_g (x)$.
Then, remembering that the nonforward nucleon matrix elements of
$T_q^{++}$ and $T_g^{++}$ are parametrized as
\begin{eqnarray}
 \langle P + \frac{\Delta}{2}, s \,| \,T^{++}_{q/g} (0) \,| 
 \,P - \frac{\Delta}{2}, s \rangle
 &=& A_{q/g} (\Delta^2) \,P^+ \,P^+ \ + \ 
 \frac{B_{q/g}(\Delta^2)}{M} \,P^+ \,\epsilon^{+ \alpha \beta \sigma} \,
 s_\alpha \,P_\beta \,\,i \,\Delta_\sigma \nonumber \\
 &+& \ C_{q/g} (\Delta^2) \,M^2 \,g^{++} \ + \ O(\Delta^2),
\end{eqnarray}
we can easily get the following sum rules :
\begin{eqnarray}
 \int_{-1}^1 \,x \,f_q (x) \,dx &=& A_q (0), \\
 \int_{-1}^1 \,x \,f_g (x) \,dx &=& A_g (0).
\end{eqnarray}
These quantities are nothing but the momentum fractions $\langle x \rangle^q$
and $\langle x \rangle^g$ carried by the quark and gluon fields in the
nucleon.
The famous momentum sum rule of QCD
\begin{equation}
 \int_{-1}^1 \,x \,[\, f_q (x) \ + \ f_g (x) \,] \,dx \ = \ 
 \langle x \rangle^q \ + \ \langle x \rangle^g \ = \ 1 ,
\end{equation}
then follows from the equation
\begin{equation}
 \langle P, s \,| \,T_q^{++} (0) \ + \ T_g^{++} (0) \,| \,P, s \rangle
 \,/ \, (P^+)^2 \ = \ 1.
\end{equation}
On the other hand, by differentiating the relations (\ref{q-gpd})
and (\ref{g-gpd}) before taking the forward limit, we obtain the
identities
\begin{eqnarray}
 \left. - \,i \,P^+ \,\frac{\partial}{\partial \Delta_\sigma} \,
 \int_{-1}^1 \,x \,f_q (x, 0, \Delta) \,dx \right|_{\Delta = 0}
 &=& \frac{B_q (0)}{M} \,\epsilon^{+ \sigma \alpha \beta} \,
 s_\alpha \,P_\beta , \\
 \left. - \,i \,P^+ \,\frac{\partial}{\partial \Delta_\sigma} \,
 \int_{-1}^1 \,x \,f_g (x, 0, \Delta) \,dx \right|_{\Delta = 0}
 &=& \frac{B_g (0)}{M} \,\epsilon^{+ \sigma \alpha \beta} \,
 s_\alpha \,P_\beta .
\end{eqnarray}
Here the quantities $B_q (0)$ and $B_g (0)$ are the forward limits of
the form factors appearing in the nonforward nucleon matrix element
of quark and gluon parts of the QCD energy momentum tensor.
The fact that they are just proportional to the total angular momenta
of quark and gluon such that (see (\ref{Nspin-SR})-(\ref{Nspin-SR4}))
\begin{eqnarray}
 J_q &=& \frac{1}{2} \,B_q (0), \\
 J_g &=& \frac{1}{2} \,B_g (0),
\end{eqnarray}
is the famous Ji sum rule \cite{Ji97PRL},\cite{Ji97}.
To avoid confusion, we recall here that
the above form factors $B_{q/g} (\Delta^2)$ are related to more familiar
form factors $A^{q/g}_{20} (\Delta^2)$ and $B^{q/g}_{20} (\Delta^2)$
through the relation $B_{q/g} (\Delta^2) = A^{q/g}_{20} (\Delta^2) + 
B^{q/g}_{20} (\Delta^2)$. Here, $A^{q/g}_{20} (\Delta^2)$ and
$B^{q/g}_{20} (\Delta^2)$ are respectively the 2nd moments of the
unpolarized GPDs $H^{q/g} (x, \xi, \Delta^2)$ and $E^{q/g} (x, \xi, \Delta^2)$
with $\xi = 0$, so that
\begin{eqnarray}
 B_{q/g} (\Delta^2) &=& A_{20}^{q/g} (\Delta^2) \ + \ B_{20}^{q/g} (\Delta^2)
 \nonumber \\
 &=& \int_{-1}^1 \,x \,\left[H^{q/g} (x,0,\Delta^2) \ + \ 
 E^{q/g} (x,0,\Delta^2) \, \right] \,dx .
\end{eqnarray}

The GPDs are measurable quantities so that
$J_q$ and $J_g$ can in principle be determined empirically.
Once $J_q$ and $J_g$ are known, it is clear from our general formula
for the nucleon spin decomposition that the orbital angular momenta
$L_q$ and $L_g$ of the quarks and gluons can be extracted just by
subtracting the intrinsic spin parts of the quarks and gluons, i.e.
$\frac{1}{2} \,\Delta q$ and $\Delta g$. A remaining critical question is
then as follows. Can the intrinsic quark and gluon spin parts defined
in our gauge-invariant decomposition of the nucleon spin be identified
with the corresponding quantities as measured by the high-energy
DIS measurements ?
This is a fairly delicate question especially for the gluon polarization
$\Delta g$. However, the importance of this question should not be dismissed.
In fact, only in the case we could affirmatively answer this
question, we would attain a sound theoretical basis for a
completely meaningful gauge-invariant decomposition of the nucleon spin.

To answer the raised question, it is useful to remember the
investigation by Bashinsky and Jaffe \cite{BJ99}, which can be thought of as a
nontrivial generalization of the light-cone-gauge formulation of
parton distribution functions.
The reason why we pay special attention to the formulation of Bashinky and
Jaffe is twofold.
The first reason is of course that their light-cone-gauge formulation
of the parton distribution functions and the corresponding 1st moments
just fits our program, which aims at finding the relation between the
gluon spin term in our decomposition and high-energy deep-inelastic-scattering
observables. Another important reason, although not unrelated to the first,
is that we want to show explicitly the fact that the numerical value of the
gluon spin term in the Bashinsky-Jaffe decomposition just coincides with
that of the gluon spin term of our more general decomposition.
(To avoid confusion, however, we emphasize once again that the orbital angular
momentum parts of quark and gluons in the Bashinsky-Jaffe decomposition are
never related to the corresponding terms in our recommendable decomposition (I)
by any gauge transformation. See the discussion later for more detail.)

Starting with the standard light-cone-gauge formulation of
parton distribution functions, Bashinsky and Jaffe invented
a method of constructing gauge-invariant quark and gluon distributions
describing abstract QCD observables and apply this formalism for
analyzing angular momentum contents of the nucleon. In addition to the
known quark and gluon polarized distribution functions, they gave a
definition of gauge-invariant distributions for quark and gluon orbital
angular momentum. According to their notation, these distribution functions
for the quark and gluon spin and orbital angular momenta are given by
\begin{eqnarray}
 f_{\Delta q} (x_{Bj}) &=& \frac{1}{2 \,\pi \,\sqrt{2}} \,
 \int \,d \xi^- \,e^{\,i \,x_{Bj} \,P^+ \,\xi^-} \,
 \langle P \,| \,\psi^\dagger_+ (0) \,\gamma^5 \,\psi_+ (\xi^-) 
 \,P \rangle , \\
 f_{L_q} (x_{Bj}) &=& 
 \frac{\int \,d \xi^- \,e^{\,i \,x_{Bj} \,P^+ \,\xi^-} \,
 \langle P \,| \,\int \,d^2 x^\perp \,\psi^\dagger_+ (x^\perp)\,
 \left(\, x^1 \,i \,{\cal D}_2 - x^2 \,i \,{\cal D}_1 \,) \,
 \psi_+ (x^\perp + \xi^-) \,|\, P \right\rangle}
 {2 \,\pi \,\sqrt{2} \,\left(\,\int \,d^2 x^\perp \,\right)} , \
 \ \ \ \ \ \\
 f_{\Delta g} (x_{Bj}) &=& \frac{1}{4 \,\pi} \,
 \int \,d \xi^- \,e^{\,i \,x_{Bj} \,P^+ \,\xi^-} \,
 \langle P \,| \,F^{+ \lambda} (0) \,
 \epsilon^{+-}{}_\lambda{}^\chi \,
 A_\chi (\xi^-) \,|\,P \rangle , \\
 f_{L_g} (x_{Bj}) &=& 
 \frac{i \,\int \,d \xi^- \,e^{\,i \,x_{Bj} \,P^+ \,\xi^-} \,
 \langle P \,| \,\int \,d^2 x^\perp \,F^{+ \lambda} (x^\perp)\,
 \left(\, x^1 \,i \,{\cal D}_2 - x^2 \,i \,{\cal D}_1 \,) \,
 A_\lambda (x^\perp + \xi^-) \,|\, P \right\rangle}
 {4 \,\pi \,\left(\,\int \,d^2 x^\perp \,\right)} .
 \ \ \ \ \ \ \ 
\end{eqnarray}
Here, $\psi_+ \equiv \frac{1}{2} \,\gamma^- \,\gamma^+ \,\psi$, and
\begin{equation}
 {\cal D}_i \ = \ \partial_i \ - \ i \,g \,{\cal A}_i ,
\end{equation}
denotes the residual gauge covariant derivative, corresponding to the
residual gauge degrees of freedom remaining after taking the light-cone
gauge $A^+ = 0$. The 1st moments of these distribution functions
becomes
\begin{eqnarray}
 \Delta q &=& \frac{1}{\sqrt{2} \,P^+} \,\langle P \,| \,
 \psi^\dagger_+ (0) \,\gamma^5 \,\psi_+ (0) \,| \,P \rangle , \\
 L_q &=& \frac{1}{\sqrt{2} \,P^+ \,\left(\,\int \,d^2 x^\perp \,\right)} \,
 \langle P \,| \,\int \,d^2 x^\perp \,\psi^\dagger_+ (x^\perp) \,
 (\,x^1 \,i \,{\cal D}_2 \ - \,x^2 \,i \,{\cal D}_1 \,) \,
 \psi_+ (x^\perp) \,| \,P \rangle , \ \ \ \\
 \Delta g &=& \frac{1}{2 \,P^+} \,\langle P \,| \,
 F^{+ \lambda} (0) \,\epsilon^{+-}{}_\lambda{}^\chi \,
 A_\chi (0) \,|\,P \rangle , \\ \label{delta-g-bj}
 L_g &=& \frac{1}{2 \,P^+ \,\left(\,\int \,d^2 x^\perp \,\right)} \,
 \langle P \,| \,\int \,d^2 x^\perp \,F^{+ \lambda} (x^\perp)\,
 (\, x^1 \,i \,{\cal D}_2 - x^2 \,i \,{\cal D}_1 \,) \,
 A_\lambda (x^\perp) \,| \,P \rangle .
\end{eqnarray}
One might notice here the resemblance of this decomposition to our
decomposition (II).
To see it more closely, we take the nucleon matrix
element of $M^{\prime \mu \nu \lambda}$ in (\ref{decomposition2})
with $\mu = +, \nu = 1, \lambda = 2$ :
\begin{eqnarray}
 \langle P, s \,| \,M^{\prime + 1 2} (0) \,
 | \,P, s \rangle 
 &=& 
 \langle P, s \,| \,M^{\prime + 1 2}_{q-spin} (0) \,
 | \,P, s \rangle
 \ + \ 
 \langle P, s \,| \,M^{\prime + 1 2}_{q-OAM} (0) \,
 | \,P, s \rangle \nonumber \\
 &+&
 \langle P, s \,| \,M^{\prime + 1 2}_{g-spin} (0) \,
 | \,P, s \rangle
 \ + \ 
 \langle P, s \,| \,M^{\prime + 1 2}_{g-OAM} (0) \,
 | \,P, s \rangle , 
\end{eqnarray}
where
\begin{eqnarray}
 M^{\prime + 1 2}_{q-spin} &=& 
 \frac{1}{2} \,\bar{\psi} \,\gamma_3 \,\gamma_5 \,\psi \ = \ 
 \psi^\dagger_+ \,\gamma_5 \,\psi_+ , \\
 M^{\prime + 1 2}_{q-spin} &=&
 \bar{\psi} \,\gamma^+ \,
 (\,x^1 \,i \,D^2_{pure} \ - \ x^2 \,i \,D^1_{pure} \,) \,\psi \ = \ 
 2 \,\psi^\dagger_+ \,
 (\,x^1 \,i \,D^2_{pure} \ - \ x^2 \,i \,D^1_{pure} \,) \,\psi_+ ,
 \ \ \ \ \ \\
 M^{\prime + 1 2}_{g-spin} &=& 
 2 \,\mbox{Tr} \,[\,F^{+ 2} \,A^1_{phys} \ - \ F^{+ 1} \,A^2_{phys} \,]
 \ = \ 2 \,\mbox{Tr} \,[\,F^{+ \lambda} \,
 \epsilon^{+-}{}_\lambda{}^\chi \,A^{phys}_\chi \,] , \\ \label{delta-g}
 M^{\prime + 1 2}_{g-OAM} &=& 
 - 2 \,\mbox{Tr} \,[\,F^{+ \lambda} \,
 (\,x^1 \,D^2_{pure} \ - \ x^2 \,D^1_{pure} \,) \,A^{phys}_\lambda \,] . 
\end{eqnarray}
Here, we have omitted the boosts and total derivative terms, which are
irrelevant in our discussion here.

The above perfect correspondence indicates the following. The residual
gauge covariant derivative ${\cal D}_i = \partial_i - i \,g \,{\cal A}_i$
appearing in the orbital parts of the Bashinsky-Jaffe decomposition
is critically different from the standard covariant derivative containing the
full gauge field. The field ${\cal A}_i$ contained in ${\cal D}_i$ would rather
correspond to the pure-gauge part $A^{pure}_i$ in our general framework.
(This fact will soon be confirmed in more explicit form.)
This means that the quark and gluon orbital angular momenta appearing
in the Bashinsky-Jaffe decomposition are basically the canonical ones not
the dynamical ones. In fact, we have already pointed out in sect.II that
the Bashinsky-Jaffe decomposition and the Chen et al. decomposition fall
into the same category in the sense that they are both nontrivial gauge-invariant
extensions of the gauge-variant Jaffe-Manohar decomposition.
As repeatedly emphasized, this is not our recommendable decomposition,
since no practical experimental process is known for measuring the above
distribution functions for the quark and gluon orbital angular momenta and
the corresponding 1st moments.

Despite this fact, one should clearly recognize the fact that the quark and
gluon spin terms in the decomposition (II) are exactly the same as those of our
recommendable decomposition (I), i.e.
\begin{eqnarray}
 M^{\prime \mu \nu \lambda}_{q-spin} &=& M^{\mu \nu \lambda}_{q-spin} ,\\
 M^{\prime \mu \nu \lambda}_{q-spin} &=& M^{\mu \nu \lambda}_{q-spin} .
\end{eqnarray}
We therefore concentrate on the relationship between the quark and gluon
spin terms in the Bashinsky-Jaffe decomposition and those of our
gauge-invariant decomposition (I).
There is no problem with the quark spin
part. In fact, this term is trivially gauge-invariant in itself and
it has been long known that it can be measured through polarized
DIS measurements. The quark spin term in our decomposition precisely
coincide with this measurable quantity.

The gluon spin part is a little more delicate, however. In fact, it is
often claimed that there is no gauge-invariant decomposition of gluon
total angular momentum into its spin and orbital parts. Since the
fundamental gauge principle dictates that observables must be
gauge-invariant, one might suspect whether $\Delta g$ is really an
observable quantity or not. To clear up these unsettled issues,
we first recall that, in our gauge-invariant decomposition of the covariant
angular momentum tensor, we do not actually need to fix gauge explicitly.
Only conditions necessary in our decomposition is that $A^\mu_{pure}$
in $A^\mu = A^\mu_{phys} + A^\mu_{pure}$ satisfies the pure-gauge
requirement, $F^{\mu \nu}_{pure} \equiv \partial^\mu \,A^\nu_{pure} - 
\partial^\nu \,A^\mu_{pure} - i \,g \,[A^\mu_{pure}, A^\nu_{pure}]$
and the appropriate gauge transformation properties (\ref{gtr-phys})
and (\ref{gtr-pure}) of $A^\mu_{phys}$
and $A^\mu_{pure}$. (The fact is that $A^\mu_{phys}$ basically contains
only the gauge-independent and physics-containing part common to  
all gauges that recides on the {\it phsical} plane \cite{Wong10}.)

Now, assume that we impose the light-cone gauge condition $A^+ = 0$,
while leaving the freedom of residual gauge transformation
retaining $A^+ = 0$. Comparing (\ref{delta-g-bj}) and
(\ref{delta-g}), it must be clear by now that the gluon spin terms in the
Bashinsky-Jaffe decomposition can be thought of as the ``light-cone-gauge
fixed form'' of our more general expression. However, careful readers
might notice a delicate difference between the two expressions
(\ref{delta-g-bj}) and (\ref{delta-g}).
In the gluon spin term in our decomposition, what enters is
$A^{phys}_\chi$, i.e. the physical part of $A_\chi$, whereas the full
gauge field $A_\chi$ enters in the $\Delta g$ term of the
Bashinsky-Jaffe decomposition. As such, the fully gauge-invariant nature
of the $\Delta g$ term in the Bashinsky-Jaffe decomposition is not so
obvious, which is a source of confusion. Now we will show that the full
gauge field $A_\chi$ in this $\Delta g$ term can be replaced by
its physical part $A^{phys}_\chi$ without any approximation.
(Although not so clearly written, this fact was already recognized in the
paper by Bashinsky and Jaffe \cite{BJ99}.)

The proof goes as follows. Following Bashinsky and Jaffe \cite{BJ99},
we introduce the Fourier decomposition of
$A_\lambda (\xi) \equiv A^{LC}_\lambda (\xi)$ as
\begin{equation}
 A_\lambda (\xi) \ = \ \int \,\frac{d k^+}{2 \,\pi} \,
 e^{\,- \,i \,k^+ \,\xi^-} \,\tilde{A}_\lambda (k^+, \tilde{\xi}),
\end{equation}
where
\begin{equation}
 \tilde{\xi} \ = \ (\xi^+, \xi^1, \xi^2) \ = \ (\xi^+, \xi^\perp).
\end{equation}
There still remains a residual gauge symmetry. In fact, the condition
$A^+ = 0$ is preserved by a gauge transformation, the parameters
of which do not depend on the coordinate $\xi^-$.
Under such gauge transformation,
$\tilde{A}_\lambda (k^+, \tilde{\xi})$ transforms as
\begin{equation}
 \tilde{A}_\lambda (k^+ , \tilde{\xi}) \ \rightarrow \ 
 U(\tilde{\xi}) \,\left(\,\tilde{A}_\lambda (k^+, \tilde{\xi}) \ + \ 
 \frac{2 \,\pi \,i \,\delta(k^+)}{g} \,\,\partial_\lambda \,\right) \,
 U^{-1} (\tilde{\xi}).
\end{equation}
Here the inhomogeneous term appears only at $k^+ = 0$. This motivates
them to split the fields $\tilde{A}_\lambda (k^+, \tilde{\xi})$ into
two parts as
\begin{equation}
 \tilde{A}_\lambda (k^+, \tilde{\xi}) \ = \ 
 2 \,\pi \,\delta (k^+) \, {\cal A}_\lambda (\tilde{\xi}) \ + \ 
 \tilde{G}_\lambda (k^+, \tilde{\xi}) ,
\end{equation}
which respectively transform as
\begin{eqnarray}
 \tilde{G}_\lambda (k^+, \tilde{\xi}) &\rightarrow& 
 U (\tilde{\xi}) \,\tilde{G}_\lambda (k^+, \tilde{\xi}) \,
 U^{-1} (\tilde{\xi}), \\
 {\cal A}_\lambda (\tilde{\xi}) &\rightarrow& \ U (\tilde{\xi}) \,
 \left(\,{\cal A}_\lambda (\tilde{\xi}) \ + \ \frac{i}{g} \,\,
 \partial_\lambda \,\right) \,U^{-1} (\tilde{\xi}) ,
\end{eqnarray}
under the residual gauge transformation that does not depend on $\xi^-$.
The decomposition is unique if one requires the boundary condition
$\tilde{G}_\lambda (k^+, \tilde{\xi}) |_{k^+ = 0} \equiv 0$ \cite{BJ99}.
In the coordinate space, this corresponds to the decomposition
\begin{equation}
 A_\lambda (\xi) \ = \ A_\lambda^{phys} (\xi) \ + \ 
 A_\lambda^{pure} (\xi) ,
\end{equation}
with
\begin{eqnarray}
 A_\lambda^{phys} (\xi) &\equiv& \int \,\frac{d k^+}{2 \,\pi} \,
 e^{\ - \,i \,k^+ \,\xi^-} \,\tilde{G}_\lambda (k^+,\tilde{\xi}), \\
 A_\lambda^{pure} (\xi) &\equiv& \int \,\frac{d k^+}{2 \,\pi} \,
 e^{\ - \,i \,k^+ \,\xi^-} \,\,2 \,\pi \,\delta (k^+) \,
 {\cal A}_\lambda (\tilde{\xi}) \ = \ {\cal A}_\lambda (\tilde{\xi})
 \ = \ {\cal A}_\lambda (\xi^+, \xi^\perp) .
\end{eqnarray}
A noteworthy fact here is that the pure gauge part of $A_\lambda (\xi)$
does not depend on the coordinate $\xi^-$. By making use of it,
one can easily convince that these two parts transform in the following way
\begin{eqnarray}
 A_\lambda^{phys} (\xi) &\rightarrow& U (\tilde{\xi}) \,
 A_\lambda^{phys} (\xi) \,U^{-1} (\tilde{\xi}) , \\
 A_\lambda^{pure} (\xi) &\rightarrow& U (\tilde{\xi}) \,
 \left(\,A_\lambda^{pure} (\xi) \ + \ \frac{i}{g} \,\,
 \partial_\lambda \,\right) \,U^{-1} (\tilde{\xi}) ,
\end{eqnarray}
under the residual gauge transformation.
This transformation rules just confirm our previous statement on the
correspondence
\begin{eqnarray*}
 {\cal A}_\lambda \ &\longleftrightarrow& \ A^{pure}_\lambda, \\
 {\cal D}_i \ = \ \partial_i - g \,{\cal A}_i \ &\longleftrightarrow& \  
 D^{pure}_i \ = \ \partial_i - g \,A^{pure}_i . 
\end{eqnarray*}
More precisely, ${\cal A}_\lambda$ can be thought of as a special case of
our more general quantity $A^{pure}_\lambda$ after choosing the
light-cone gauge. This implies that the gluon spin term in our general
(gauge-invariant) decomposition in fact reduces to the corresponding
piece of the Bashinsky-Jaffe decomposition given in the light-cone gauge.

Now we return to the expression
for the polarized gluon distribution function $f_{\Delta g} (x)$.
\begin{equation}
 f_{\Delta g} (x) \ = \ 
 \frac{\frac{1}{4 \,\pi} \,\int \,d^2 \xi^\perp \,
 \int \,d \xi^- \,\int \,d \eta^- \,
 e^{\,- \,i \,x \,P^+ \,\eta^-} \,\langle P \,| \,F^{+ \lambda} (\xi) \,
 \epsilon^{+-}{}_\lambda{}^\chi \,A_\chi (\xi + \eta^-) \,| \,P \rangle}
 {\int \,d^2 \xi^\perp \,\int \,d \xi^-} ,
\end{equation}
with
\begin{equation}
 \xi \ = \ (\xi^+, \xi^-, \xi^\perp), \ \ \ \tilde{\xi} 
 \ = \ (\xi^+, \xi^\perp).
\end{equation}
Noting the fact that $A_\chi^{pure} (\xi) = {\cal A}_\chi (\tilde{\xi})$
does not depend on $\xi^-$, the contribution of the pure gauge part is
given by
\begin{equation}
 f_{\Delta g} (x) \ = \ 
 \frac{\frac{\delta (x)}{2 \,P^+} \,\int \,d^2 \xi^\perp \,\int \,d \xi^- \,
 \langle P \,| \,F^{+ \lambda}(\xi) \,
 \epsilon^{+-}{}_\lambda{}^\chi \,\tilde{A}_\chi (\tilde{\xi}) \,
 | \,P \rangle}{\int \,d^2 \xi^\perp \,\int \,d \xi^-} .
\end{equation}
Using the relation $F^{+ \lambda} (\xi) = \frac{\partial}{\partial \xi^-} \,
A^\lambda (\xi)$ that hold in the light-cone gauge, we therefore find that
\begin{eqnarray}
 &\,& \int \,d \xi^- \,\langle P \,| \,F^{+ \lambda} (\xi) \,\,
 \epsilon^{+-}{}_\lambda{}^\chi \,\, {\cal A} (\tilde{\xi}) \,| \,P \rangle
 \nonumber \\
 &=& 
 \int \,d \xi^- \,\langle P \,| \,\frac{\partial}{\partial \xi^-} \,
 A^\lambda (\xi) \,\,
 \epsilon^{+-}{}_\lambda{}^\chi \,\, {\cal A} (\tilde{\xi}) \,| \,P \rangle
 \ \ \ \ \nonumber \\
 &=& \langle P \,| \,\left[\,A^\lambda (\xi^- = + \,\infty) - 
 A^\lambda (\xi^- = - \,\infty) \,\right] \,
 \epsilon^{+-}{}_\lambda {}^\chi \,{\cal A}_\chi (\tilde{\xi}) \,| \,
 P \rangle \nonumber \\
 &-&  
 \int \,d \xi^- \,\langle P \,| \,A^\lambda (\xi) \,\,
 \epsilon^{+-}{}_\lambda{}^\chi \,\,\frac{\partial}{\partial \xi^-} \,
 {\cal A}_\chi (\tilde{\xi}) \,| \,P \rangle \nonumber \\
 &=&
 \langle P \,| \,\left[\,A^\lambda (\xi^- = + \,\infty) - 
 A^\lambda (\xi^- = - \,\infty) \,\right] \,
 \epsilon^{+-}{}_\lambda {}^\chi \,{\cal A}_\chi (\tilde{\xi}) \,| \,
 P \rangle ,
\end{eqnarray}
since $\frac{\partial}{\partial \xi^-} \,{\cal A}_\chi (\tilde{\xi}) = 0$.
In the light-cone gauge, the above surface term does not vanish, because
either of $A^\lambda (\xi^- = + \,\infty)$ or
$A^\lambda (\xi^- = - \,\infty)$ or both remains finite.
Nonetheless, as pointed out in \cite{BJ99}, the surface term does
not contribute to the polarized gluon distribution $f_{\Delta G} (x)$, since
\begin{eqnarray}
 &\,& \int_{- \,\infty}^{+ \,\infty} \,d \xi^- \,
 \langle P \,| \,F^{+ \lambda} (\xi) \,\,
 \epsilon^{+-}{}_\lambda {}^\chi \,\,
 {\cal A}_\chi (\tilde{\xi}) \,P \rangle \,/ \, 
 \int_{- \,\infty}^{+ \,\infty} \,d \xi^- \nonumber \\
 &=& \langle P \,| \,\left[\,A^\lambda (\xi^- = + \,\infty) - 
 A^\lambda (\xi^- = - \,\infty) \,\right] \,
 \epsilon^{+-}{}_\lambda {}^\chi \,\, {\cal A}_\chi (\tilde{\xi}) \,| \,
 P \rangle \,/ \,
 \int_{- \,\infty}^{+ \,\infty} \,d \xi^- \nonumber \\
 &=& \ \ 0 .
\end{eqnarray}

On the other hand, the contribution of the physical part of
${\cal A}_\chi$ is given as
\begin{equation}
 f_{\Delta g} (x) \ = \ \frac{\frac{1}{4 \,\pi} \,\int \,d^2 \xi^\perp \,
 \int \,d \xi^- \,\int \,d \eta^- \,e^{\,i \,x \,P^+ \,\eta^2} \,
 \langle P \,| \,F^{+ \lambda} (\xi) \,\epsilon^{+-}{}_\lambda{}^\chi \,
 A_\chi^{phys} (\xi + \eta^-) \,| \,P \rangle}
 {\int \,d^2 \xi^\perp \,\int \,d \xi^-} .
\end{equation}
Using the translational invariance
\begin{eqnarray}
 \langle P \,| \,F^{+ \lambda} (\xi) \, \epsilon^{+-}{}_\lambda{}^\chi \,
 A_\chi^{phys} (\xi + \eta^-) \,| \,P \rangle \ = \ 
 \langle P \,| \,F^{+ \lambda} (0) \, \epsilon^{+-}{}_\lambda{}^\chi \,
 A_\chi^{phys} (\eta^-) \,| \,P \rangle , 
\end{eqnarray}
we therefore obtain
\begin{equation}
 f_{\Delta g}^{phys} (x) \ = \ \frac{1}{4 \,\pi} \,\int \,d \eta^- \,
 e^{\,i \,x \,P^+ \,\eta^-} \,\langle P \,| \,F^{+ \lambda} (0) \,
 \epsilon^{+-}{}_\lambda{}^\chi \,A_\chi^{phys} (\eta^-) \,| \,P \rangle .
\end{equation}
The corresponding 1st moment becomes
\begin{equation}
 \Delta g \ = \ \int_{-1}^1 \,f_{\Delta G}^{phys} (x) \,dx \ = \ 
 \frac{1}{2 \,P^+} \,\langle P \,| \,F^{+ \lambda} (0) \,
 \epsilon^{+-}{}_\lambda{}^\chi \,A_\chi^{phys} (0) \,| \,P \rangle .
 \label{Delta-g-BJ}
\end{equation}
Note that this precisely takes the same form as our $\Delta g$ term
\begin{eqnarray}
 \Delta g &=& \langle P, s \,| \,
 M_{g-spin}^{+12} (0) \,P, s \rangle \,/ \,2 \,P^+ 
 \ = \ 
 \frac{1}{2 \,P^+} \,\langle P \,| \,F^{+ \lambda} (0) \,
 \epsilon^{+-}{}_\lambda{}^\chi \,A_\chi^{phys} (0) \,| \,P \rangle .
 \label{Delta-g-I}
\end{eqnarray}
Needless to say, $A^{phys}_\chi$ in (\ref{Delta-g-BJ}) should be
interpreted as a gauge fixed-form of more general $A^{phys}_\chi$ in
(\ref{Delta-g-I}) after taking the light-cone gauge.
With this understanding, it is clear now that the numerical value of the
gluon spin term in the Bashinsky-Jaffe decomposition precisely
coincides with that of the gluon spin term in our more general
decomposition. (This is just what is meant by the gauge-invariance !) 
To put it in another way, the gluon spin part in our
gauge-invariant decomposition precisely reduces
to the 1st moment of the polarized
gluon distribution accessed by high-energy DIS measurements.
It is widely recognized that there is no gauge-invariant
local operator corresponding to the 1st moment of the polarized gluon
distribution in the standard
operator-product expansion. However, it should be clear by now that
there is no conflict between this general statement
and our finding above. The decomposition of the gauge field $A_\mu$ into
its physical and pure-gauge parts is generally a nonlocal operation so that
$A_\chi^{phys}$ is not a local operator. (This is true not only for the
light-cone gauge but also for the generalized Coulomb gauge advocated by
Chen et al.) Now we can definitely say that the gluon spin contribution to
the nucleon spin measured by high-energy DIS measurements just coincide with
the quantity appearing in our general decomposition of the nucleon spin
discussed in the previous sections, so that it can be given a manifestly
gauge-invariant and practically frame-independent meaning.

At this point, it may be useful to summarize some of the important lessons,
which we have learned from the present investigation.
First, as repeatedly emphasized, the way of gauge invariant decomposition of
the nucleon spin is not necessarily unique. We showed that there are basically
two independent decompositions of the nucleon spin, i.e. the decomposition (I)
specified by (\ref{decomposition1}) and the decomposition (II) specified by
(\ref{decomposition2}). The decomposition (II) contains three known
decompositions of the nucleon spin, i.e. those of Jaffe and Monahar, of
Bashinsky and Jaffe, and of Chen et al. We can say that all these decompositions
are physically equivalent in the sense that they are all obtained from more
general decomposition (II) by means of suitable gauge-fixing.
On the other hand, the physical content of the decomposition (I)
is critically different from the decomposition (II). The decomposition (I)
contains the famous Ji decomposition, although the former allows the
decomposition of the total gluon angular momentum into its intrinsic spin and
orbital parts, which was given up in the latter.
For pedagogical reason, we think it useful to summarize this state of affairs
in a conceptual figures as illustrated in Fig.1.
\vspace{3mm}
\begin{figure}[ht]
\begin{center}
\includegraphics[width=12cm]{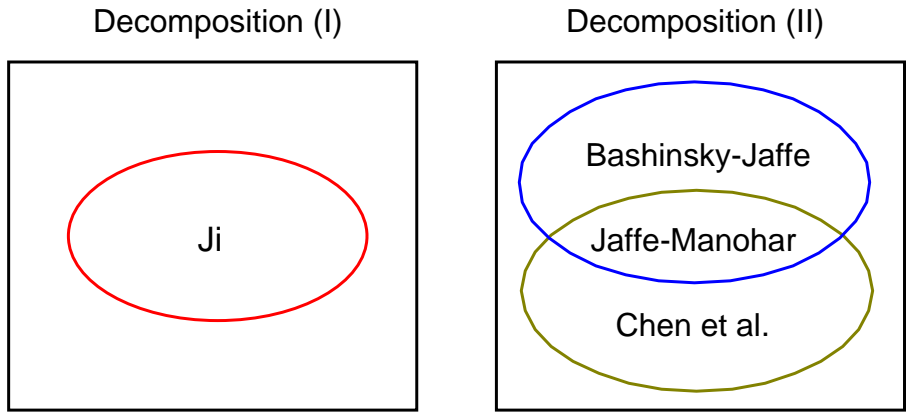}
\caption{\label{label}Schematic picture of two independent gauge-invariant
decompositions of nucleon spin and the relation with the known decompositions.}
\end{center} 
\end{figure}

The superiority of the decomposition (I) over the decomposition (II)
is that both of the quark and gluon orbital angular momenta can be related to
concrete high-energy observables.
In fact, after confirmation of the frame-independence of our nucleon spin
decomposition, we can now work in an arbitrary Lorentz frame.
Then, the following identity must hold for the quark orbital angular momentum
in the decomposition (I) : 
\begin{eqnarray}
 L_q &=& \langle p \uparrow \,| \,M^{012}_{q-OAM} \,| p \uparrow \rangle \nonumber \\
 &=& \frac{1}{2} \,\int_{-1}^1 \,x \,\left[\, H^q (x,0,0) \ + \ E^q (x,0,0) \,\right] \,dx
 \ + \ \frac{1}{2} \,\int_{-1}^1 \,\Delta q(x) \,dx , \label{Lq_GPDPDF}
\end{eqnarray}
where
\begin{equation}
 M^{012}_{q-OAM} \ = \ \bar{\psi} \,\left(\,\bm{x} \times
 \frac{1}{i} \,\bm{D} \,\right)^3 \,\psi.
\end{equation}
(Note that the GPDs and the polarized PDF appearing in the r.h.s. of
(\ref{Lq_GPDPDF}) are the Lorentz-frame independent quantities.)
This identity means that the quark orbital angular momentum $L_q$ in the
decomposition (I) precisely coincides with the difference of the 2nd
moment of the unpolarized GPD $H^q (x,0,0) + E^q (x,0,0)$ and the 1st
moment of the longitudinally polarized quark distribution $\Delta q(x)$,
which are both observables. Furthermore, this $L_q$ is given as a proton
matrix element of the {\it dynamical} orbital angular momentum of quarks, i.e. 
$\bm{x} \times \frac{1}{i} \,\bm{D} = \bm{x} \times \frac{1}{i} \,
\left( \nabla - i \,g \,\bm{A} \,\right)$ not the {\it canonical} orbital angular
momentum $\bm{x} \times \frac{1}{i} \,\nabla$ or its gauge-invariant
extension $\bm{x} \times \frac{1}{i} \,\bm{D}_{pure} = \bm{x} \times \,\frac{1}{i} \,
\left( \nabla - i \,g \,\bm{A}_{pure} \,\right)$ \cite{Ji97}.

Similarly, for the gluon orbital angular momentum $L_g$ in the decomposition (I),
the following identity must hold
\begin{eqnarray}
 L_g &=& \langle p \uparrow \,| \,M^{012}_{g-OAM} \,| p \uparrow \rangle \nonumber \\
 &=& \frac{1}{2} \,\int_{-1}^1 \,x \,\left[\, H^g (x,0,0) \ + \ E^g (x,0,0) \,\right] \,dx
 \ + \ \int_{-1}^1 \,\Delta g(x) \,dx ,
\end{eqnarray}
where
\begin{eqnarray}
 M^{012}_{g-OAM} &=& 2 \,\mbox{Tr} \,\left[\, E^j \,
 \left(\,\bm{x} \times \bm{D}_{pure} \,\right)^3 \,A^{phys}_j \,\right] 
 \ + \ 2 \,\mbox{Tr} \,\left[\,\rho \,\left( \bm{x} \times \bm{A}_{phys} \,
 \right)^3 \,\right] .
 \label{g-OAM_operator}
\end{eqnarray}
One confirms that the gluon orbital angular momentum in the decomposition (I)
just coincides with the difference of the 2nd moment of the gluon
GPD $H^g (x,0,0) + E^g (x,0,0)$ and the 1st moment of the longitudinally
polarized gluon distribution $\Delta g(x)$.
What is noteworthy here is that the relevant gluon orbital angular momentum
operator entering in this identity consists of two terms.
The 1st piece is a gauge-invariant extension of the canonical orbital angular
momentum of gluons. (It is physically equivalent to the usual canonical
orbital angular momentum appearing, for instance, in the Jaffe-Manohar decomposition.)
The 2nd piece is nothing but the potential angular momentum term discussed
in some detail in our previous paper \cite{Wakamatsu10}.
In view of the analogous situation for the quark part, it would be legitimate now
to call the sum of these two pieces, i.e. the whole
part of $M^{012}_{g-OAM}$ in (\ref{g-OAM_operator}), the {\it dynamical} orbital
angular momentum of gluon field. 

Before ending this section, we think it instructive to call attention
to some other recent investigations related to the nucleon spin
decomposition.
As emphasized above, the quark orbital angular momentum
extracted from the combined analysis of the unpolarized GPDs and
the longitudinally polarized quark distribution
functions is the {\it dynamical} orbital angular momentum not the
{\it canonical} one or its nontrivial gauge-invariant extension.
At least until now, we have had no means to extract the canonical
orbital angular momentum purely experimentally, which also means that
the difference between the dynamical and canonical
orbital angular momenta is not a direct experimental observable.
Nevertheless, it is not impossible to estimate the size of this
difference within the framework of a certain model.
In fact, Burkardt and BC estimated the difference between the orbital
angular momentum obtained from the Jaffe-Manohar decomposition
and that obtained from the Ji decomposition within two simple toy models,
and emphasize the possible importance of the vector potential in the
definition of orbital angular momentum \cite{BBC09}.
The difference between the above
two orbital angular momenta is nothing but the {\it potential angular
momentum} in our terminology.

Also noteworthy is recent phenomenological investigations on the role
of orbital angular momenta in the nucleon spin. In a recent paper,
we have pointed out possible existence of significant discrepancy
between the lattice QCD predictions \cite{LHPC08},\cite{LHPC10} 
for $L^u - L^d$ (the difference
of the orbital angular momenta carried by up- and down-quarks in the
proton) and the prediction of a typical low energy model of the nucleon,
for example, the refined cloudy-bag model \cite{MT88}.
It is an open question whether
this discrepancy can be resolved by strongly scale-dependent nature
of the quantity $L^u - L^d$ especially in the low $Q^2$
domain  as claimed in \cite{Thomas08}, or whether the discrepancy has
a root (at least partially) in the existence of two kinds of quark orbital
angular momenta as indicated in \cite{Waka10},\cite{WT05}.
(See also \cite{WN06},\cite{WN08} for the detail.)

\section{Summary and conclusion}

When discussing the spin structure of the nucleon, color gauge invariance
has often been a cause of controversy.
For instance, it is known that the polarized gluon distribution in
the nucleon can be defined in terms of a nucleon matrix element of 
gauge invariant correlation function. On the other hand, one is
also aware of the fact that there is no gauge-invariant local operator
corresponding to the 1st moment of the polarized gluon distribution
in the standard operator-product expansion.
Undoubtedly, this seemingly conflicting 
observation has a common root as the familiar statement that there is
no gauge-invariant decomposition of the gluon total angular momentum into 
its spin and orbital parts.
Inspired by the recent proposal by Chen et al., we find it possible
to make a gauge-invariant decomposition of covariant angular-momentum tensor 
of QCD in an arbitrary Lorentz frame. Based on this fact, we could
show that our decomposition of the nucleon spin is not only gauge-invariant
but also practically frame-independent.  
We have also succeeded to convince that each piece of our nucleon spin 
decomposition just corresponds to the observable extracted 
through combined analyses of the GPD measurements and the polarized
DIS measurements, thereby supporting the standardly-accepted experimental
program aiming at complete decomposition of the
nucleon \cite{HERMES06A}\nocite{HERMES06B}-\cite{JLabHallA07}.
In particular, the gluon spin part of our decomposition
precisely coincides with the 1st moment of the polarized gluon distribution
function. In our theoretical framework, this gluon spin part of the
decomposition is given as a nucleon matrix element of gauge-invariant
operator. However, since this operator is generally nonlocal, there
is no conflict with the knowledge of the standard operator-product
expansion.

From a practical viewpoint, more important lesson to be learned from
our present theoretical analysis would be the {\it physical insight} into
the measurable quark and gluon orbital angular momenta
appearing in our recommendable decomposition (I).  We have
confirmed that the quark orbital angular momentum, which can be extracted
as the difference of the 2nd moment of the unpolarized quark GPD and the 1st
moment of the longitudinally polarized quark distribution, is the {\it dynamical}
quark orbital angular momentum
$\langle \,\bm{x} \times (\bm{p} - g \,\bm{A}) \,\rangle$
not the {\it canonical} one $\langle \, \bm{x} \times \bm{p} \rangle$.
Similarly, the gluon orbital angular momentum extracted as the difference
of the 2nd moment of the unpolarized gluon GPD and the 1st moment of
the longitudinally polarized gluon distribution is not the {\it canonical} orbital
angular momentum but the {\it dynamical} orbital angular momentum containing
the {\it potential angular momentum} term in our terminology.
Even though no experimental process to directly access to the canonical
orbital angular momenta is known at present, one should clearly keep
in mind the existence of two kinds of orbital angular momenta for both of quarks
and gluons.

\vspace{3mm}
\begin{acknowledgments}
I would like to thank Prof.~E. Leader for critical but useful comments.
Enlightening discussion with Prof.~T. Kubota is also greatly
acknowledged. This work is supported in part by a Grant-in-Aid for
Scientific Research for Ministry of Education, Culture, Sports, Science
and Technology, Japan (No.~C-21540268)
\end{acknowledgments}


\begin{thebibliography}{10}

\bibitem{EMC88}
EMC Collaboration~: J.~Aschman~et al.,
Phys. Lett. B {\bf 206}, 364 (1988).

\bibitem{EMC89}
EMC Collaboration~: J.~Aschman~et al.,
Nucl. Phys. B {\bf 328}, 1 (1989).

\bibitem{COMPASSG06}
COMPASS Collaboration~: E.~S.~Ageev et al.,
Phys. Lett. B {\bf 633}, 25 (2006).

\bibitem{PHENIX06}
PHENIX Collaboration~: K.~Boyle et al,
AIP Conf. Proc. {\bf 842}, 351 (2006).

\bibitem{STAR06A}
STAR Collaboration~: J.~Kiryluk et al.,
AIP Conf. Proc. {\bf 842}, 327 (2006).

\bibitem{STAR06B}
STAR Collaboration~: R.~Fatemi et al.,
nucl-ex/0606007 (2006).

\bibitem{COMPASS05}
COMPASS Collaboration~: E.~S.~Ageev et al.,
Phys. Lett. B {\bf 612}, 154 (2005)

\bibitem{COMPASS07}
COMPASS Collaboration~: V.~Yu.~Alexakhin~et al.,
Phys. Lett. B {\bf 647}, 8 (2007).

\bibitem{HERMES07}
HERMES Collaboration~: A.~Airapetian~et al.,
Phys. Rev. D {\bf 75}, 012007 (2007).

\bibitem{Manohar90}
A.V.~Manohar, Phys. Rev. Lett. {\bf 65}, 2511 (1990).

\bibitem{CS82}
J.C.~Collins and D.E.~Super, Nucl. Phys. B {\bf 194}, 445 (1982).

\bibitem{JM90}
R.L.~Jaffe and A.~Manohar,
Nucl. Phys. B {\bf 337}, 509 (1990).

\bibitem{Ji97PRL}
X.~Ji, Phys. Rev. Lett. {\bf 78}, 610 (1997).

\bibitem{Wakamatsu10}
M.~Wakamatsu, Phys. Rev. D {\bf 81}, 114010 (2010).

\bibitem{Chen08}
X.~S.~Chen, X.~F.~L\"{u}, W.~M.~Sun, F.~Wang, and T.~Goldman,\\
Phys. Rev. Lett. {\bf 100}, 232002 (2008).

\bibitem{Chen09}
X.~S.~Chen, W.~M.~Sun, X.~F.~L\"{u}, F.~Wang, and T.~Goldman,\\
Phys. Rev. Lett. {\bf 103}, 062001 (2009).

\bibitem{Ji97}
X.~Ji, J. Phys. G {\bf 24}, 1181 (1998).

\bibitem{Cho10}
Y.M.~Cho, Mo-Lin Ge, and Pengming Zhang,
arXiv : 1010.1080 [nucl-th] (2010).

\bibitem{Wong10}
C.W.~Wong, Fan Wang, W.M.~Sun, and X.F.~L\"{u},
arXiv : 1010.4336 [hep-ph] (2010).

\bibitem{Konopinski78}
E.J.~Konopinski, Am. J. Phys. {\bf 46(5)}, 499 (1978).

\bibitem{BJ99}
S.V.~Bashinsky and R.L.~Jaffe, Nucl. Phys. B {\bf 536}, 303 (1999).

\bibitem{BookSakurai95}
J.J.~Sakurai, {\it Modern Quantum Mechanics}
(Addison-Wesley Pub. Co. 1995) Chap. 2.6.

\bibitem{Ji97PRD}
X.~Ji, Phys. Rev. D {\bf 58}, 056003 (1998).

\bibitem{Lubanski42}
J.K.~Lubanski, Physica (Amsterdam) {\bf 9}, 310 (1942).

\bibitem{BLT04}
B.L.G.~Bakker, E.~Leader, and T.L.~Trueman,
Phys. Rev. D {\bf 70}, 114001 (2004).

\bibitem{SW00}
G.M.~Shore and B.E.~White, Nucl. Phys. B {\bf 581}, 409 (2000).

\bibitem{AR88}
G.~Altarelli and G.G.~Ross, Phys. Lett. B {\bf 212}, 391 (1988).

\bibitem{CCM88}
R.D.~Carlitz, J.C.~Collins, and A.H.~Mueller, Phys. Lett. B {\bf 214},
229 (1988).

\bibitem{ET88}
A.V.~Efremov and O.V.~Teryaev, JINR Dubna preprint JINR EZ-88-297 (1988).

\bibitem{AEL95}
M.~Anselmino, A.~Efremov, and E.~Leader, Phys. Rep. {\bf 261}, 1 (1995).

\bibitem{LR00}
B.~Lampe and E.~Reya, Phys. Rep. {\bf 332}, 1 (2000).

\bibitem{Manohar91}
A.V.~Manohar, Phys. Rev. Lett. {\bf 66}, 1663 (1991).

\bibitem{BB91}
I.I.~Balitsky and V.M.~Braun, Phys. Lett. B {\bf 267}, 405 (1991).

\bibitem{BBC09}
M.~Burkardt and H.~BC, Phys. Rev. D {\bf 79}, 071501 (2009).

\bibitem{LHPC08}
LHPC Collaboration : Ph.~H\"{a}gler et al., Phys. Rev. D {\bf 77}, 
094502 (2008).

\bibitem{LHPC10}
LHPC Collaboration : J.D.~Bratt et al., arXiv : 1001.3620 [hep-ph].

\bibitem{MT88}
F.~Myhrer and A.W.~Thomas, Phys. Rev. D {\bf 38}, 1633 (1988).

\bibitem{Thomas08}
A.W.~Thomas, Phys. Rev. Lett. {\bf 101}, 102003 (2008).

\bibitem{Waka10}
M.~Wakamatsu,
Eur. Phys. J. A {\bf 44}, 297 (2010).

\bibitem{WT05}
M.~Wakamatsu and H.~Tsujimoto, Phys. Rev. D {\bf 71}, 074001 (2005).

\bibitem{WN06}
M.~Wakamatsu and Y.~Nakakoji,
Phys. Rev. D {\bf 74}, 054006 (2006).

\bibitem{WN08}
M.~Wakamatsu and Y.~Nakakoji,
Phys. Rev. D {\bf 77}, 074011 (2008).

\bibitem{HERMES06A}
HERMES Collaboration : Ellinghaus F {\it et al.},
Eur. J. Phys. C {\bf 46} 729 (2006).

\bibitem{HERMES06B}
HERMES Collaboration : Ye Z {\it et al.},
arXiv : hep-ex/0606061 (2006).

\bibitem{JLabHallA07}
JLab Hall A Collaboration : Mazouz M {\it et al.},
Phys. Rev. Lett. {\bf 99} 242501 (2007).

\end{thebibliography}

\end{document}